\documentclass[aps,prx,reprint,superscriptaddress]{revtex4-1}

\usepackage[T1]{fontenc} 
\usepackage{graphicx}
\usepackage{float}
\usepackage{subfigure}
\usepackage{dcolumn}
\usepackage{bm}

\usepackage{savesym}
\usepackage{amssymb}
\usepackage{amsthm}
\usepackage{amsopn}
\usepackage{booktabs}
\usepackage{multirow}
\usepackage{siunitx}
\usepackage{bbm}
\usepackage{upgreek}
\usepackage{eufrak}
\usepackage{dsfont}
\usepackage[mathscr]{eucal}
\usepackage{float}
\usepackage{braket}
\usepackage[centerlast]{caption}
\usepackage{color,soul}
\DeclareMathOperator{\Tr}{Tr}
\usepackage{latexsym}
\usepackage{amsbsy}
\usepackage[utf8]{inputenc}
\usepackage[T1]{fontenc}
\usepackage[english]{babel}

\usepackage{amsmath}

\theoremstyle{plain} 
\newtheorem{claim}{Claim}

\makeatletter
\DeclareFontFamily{OMX}{MnSymbolE}{}
\DeclareSymbolFont{MnLargeSymbols}{OMX}{MnSymbolE}{m}{n}
\SetSymbolFont{MnLargeSymbols}{bold}{OMX}{MnSymbolE}{b}{n}
\DeclareFontShape{OMX}{MnSymbolE}{m}{n}{
    <-6>  MnSymbolE5
   <6-7>  MnSymbolE6
   <7-8>  MnSymbolE7
   <8-9>  MnSymbolE8
   <9-10> MnSymbolE9
  <10-12> MnSymbolE10
  <12->   MnSymbolE12
}{}
\DeclareFontShape{OMX}{MnSymbolE}{b}{n}{
    <-6>  MnSymbolE-Bold5
   <6-7>  MnSymbolE-Bold6
   <7-8>  MnSymbolE-Bold7
   <8-9>  MnSymbolE-Bold8
   <9-10> MnSymbolE-Bold9
  <10-12> MnSymbolE-Bold10
  <12->   MnSymbolE-Bold12
}{}

\let\llangle\@undefined
\let\rrangle\@undefined
\DeclareMathDelimiter{\llangle}{\mathopen}%
                     {MnLargeSymbols}{'164}{MnLargeSymbols}{'164}
\DeclareMathDelimiter{\rrangle}{\mathclose}%
                     {MnLargeSymbols}{'171}{MnLargeSymbols}{'171}
\makeatother

\begin{document}


\title{Quantum correlations in time}


\author{Tian Zhang}
\affiliation{Department of Physics, University of Oxford, Clarendon Laboratory, Parks Road, Oxford OX1 3PU, UK}

\author{Oscar Dahlsten}
\affiliation{Institute for Quantum Science and Engineering, Department of Physics, Southern University of Science and Technology (SUSTech), Shenzhen 518055, China}
\affiliation{Department of Physics, University of Oxford, Clarendon Laboratory, Parks Road, Oxford OX1 3PU, UK}
\affiliation{London Institute for Mathematical Sciences, 35a South Street, Mayfair, London, W1K 2XF, United Kingdom}

\author{Vlatko Vedral}
\affiliation{Department of Physics, University of Oxford, Clarendon Laboratory, Parks Road, Oxford OX1 3PU, UK}
\affiliation{Centre for Quantum Technologies, National University of Singapore, 3 Science Drive 2, Singapore 117543}
\affiliation{Department of Physics, National University of Singapore, Singapore 117542}


\date{\today}

\begin{abstract}
We investigate quantum correlations in time in different approaches. We assume that temporal
correlations should be treated in an even-handed manner with spatial correlations. We compare
the pseudo-density matrix formalism with several other approaches: indefinite causal structures,
consistent histories, generalised quantum games, out-of-time-order correlations(OTOCs), and path
integrals. 
We establish close relationships among these space-time approaches 
in non-relativistic quantum theory, resulting in a unified picture. 
With the exception of amplitude-weighted correlations in the path integral formalism, 
in a given experiment, 
temporal correlations in the different approaches are the same or operationally equivalent.
\end{abstract}


\maketitle


\section{Introduction}

The problem of time~\cite{anderson2010problem} is especially notorious in quantum theory as time cannot be treated as an operator in contrast with space. 
Several attempts have been proposed to incorporate time into the quantum world in a more even-handed way to space, including: indefinite causal structures~\cite{chiribella2008quantum, chiribella2009theoretical, hardy2012operator, oreshkov2012quantum, pollock2018non, cotler2018superdensity}, consistent histories~\cite{griffiths1984consistent, griffiths2003consistent, gell2018quantum, gell1993classical, omnes1990hilbert}, generalised quantum games~\cite{buscemi2012all, rosset2018resource}, spatio-temporal correlation approches~\cite{maldacena2015bound, roberts2016chaos}, path integrals~\cite{feynman2010quantum, zinn2010path}, and pseudo-density matrices~\cite{fitzsimons2015quantum, zhao2018geometry, pisarczyk2019causal, zhang2020different}. 
Different approaches have their own advantages. Of particular interest here is the recent pseudo-density matrix approach for which one advantage is that quantum correlations in space and time are treated on an equal footing. Ref.~\cite{zhao2018geometry, zhang2020different} of  the pseudo-density matrix formalism describe how spatial and temporal correlations can be treated symmetrically in the bipartite case, for both discrete qubit systems and continuous variables. 
The present work is motivated by the need to understand how this recent approach connects to earlier approaches via temporal correlations, so that ideas and results can be transferred more readily. 

%

We accordingly aim to identify mappings between these approaches and \emph{pseudo-density matrices(PDMs)}. 
We find several mappings and relations between these approaches, including
(i) we map \emph{process matrices(PMs)} with indefinite causal order directly to pseudo-density matrices in three different ways; 
(ii) we show the diagonal terms of decoherence functionals in \emph{consistent histories(CHs)} are exactly the probabilities in temporal correlations of corresponding pseudo-density matrices; 
(iii) we show \emph{quantum-classical signalling games(QCSGs)} have the same probabilities under quantum strategies as temporal correlations measured in pseudo-density matrices; 
(iv) the number of steps in calculating \emph{out-of-time-order correlations(OTOCs)} is halved by employing by pseudo-density matrices; and 
(v) correlations in \emph{path integrals(PIs)} are defined as expectation values in terms of the amplitude measure rather than the probability measure as in pseudo-density matrices and are different from correlations in all the other approaches. 
A particular example via a tripartite pseudo-density matrix is presented to illustrate the unified picture of the different approaches. 

The paper proceeds as follows. In Section II we review classical correlations in time. We introduce the pseudo-density matrix formalism in Section III. Then we compare it with indefinite causal order in terms of forms, causality violation, quantum switch and postselection in Section IV. In Section V, we establish the relation between pseudo-density matrix and decoherence functional in consistent histories. We further explore generalised non-local games and build pseudo-density matrices from generalised signalling games in Section VI. In Section VII, we simplify the calculation of out-of-time-order correlations via pseudo-density matrices. We further provide a unified picture under a tripartite pseudo-density matrix in Section VIII. In Section IV, we argue that the path integral formalism defines correlations in a different way and does not fit into the unified picture. Finally we summarise our work and provide an outlook. 

\begin{figure}
\centering
\includegraphics[width=0.48\textwidth]{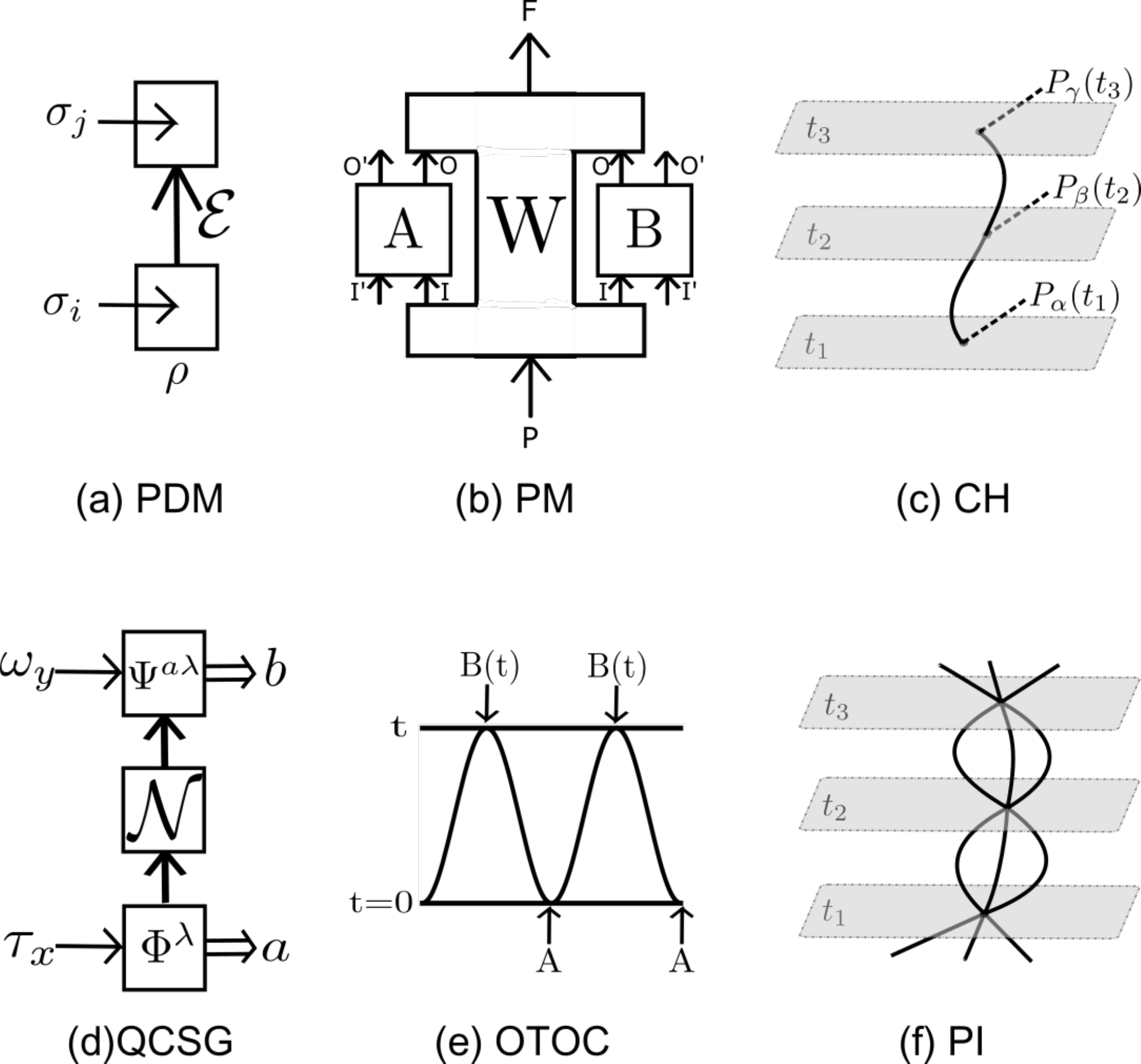}
\caption{Figure (a) represents a causally ordered pseudo-density matrix(PDM): an initial state $\rho$ evolves under the quantum channel $\mathcal{E}$ and Pauli measurement $\sigma_i$ and $\sigma_j$ are made at the initial and final time at two events separately. Figure (b) depicts a process matrix(PM) $W$ with indefinite causal structures for two parties Alice and Bob, denoted as $A$ and $B$. Both Alice and Bob have an input system $I$, an input ancillary system $I'$, an output system $O$, and an output ancillary system $O'$. The process matrix is associated with a global past $P$ and a global future $F$. Figure (c) depicts the successive measurements $P_{\alpha}(t_1) \rightarrow P_{\beta}(t_2) \rightarrow P_{\gamma}(t_3) $ at three times in consistent histories(CH). Figure (d) represents a quantum strategy of the quantum-classical signalling game(QCSG) protocol. Suppose Abby at $t_1$ receives $\tau^x_X$ and makes a measurement of instruments $\{\Phi_{X\rightarrow A}^{a|\lambda}\}$, and gains the outcome $a$. Then the quantum output goes through the quantum memory $\mathcal{N}: A \rightarrow B$. The output of the memory and $\omega^y_Y$ received by Abby at $t_2$ are fed into a measurement $\{\Psi_{BY}^{b|a, \lambda}\}$, with the outcome b. Figure (e) represents the calculation of the out-of-time order correlation function(OTOC) $\langle B(t)AB(t)A \rangle$ where $B(t)=U(t)BU^{\dag}(t)$. Figure (f) depicts three different paths go through the time slices of $t_1$, $t_2$, $t_3$ in the path integral(PI).}
\end{figure}

\section{Classical correlations in time}
We review classical correlations in time from probability theory and statistical mechanics. In the classical case, it is not necessary to distinguish spatial or temporal correlations; that is, classical correlations are defined whatever the spatio-temporal structures are. 

\subsection{Correlations in probability theory}
Now we introduce correlations defined in probability theory based on Ref.~\cite{sheldon2002first}.
For a discrete random variable $X$ with the probability mass function $p(x) = P\{X=x\}$, the expectation value of $X$ is defined as
$E[X] = \sum_{x: p(x)>0} xp(x)$.
For a continuous random variable $X$ with the probability density function $f(x)$ such that $P\{a\leq X \leq b\} = \int_a^b f(x) \textrm{d}x$, the expectation value of $X$ is defined as
$E[X] = \int_{-\infty}^{\infty} xf(x)\textrm{d}x$.
The variance of $X$ is defined as 
$\text{Var}(X) = E[(X-E[X])^2]$.
This definition is equivalent to 
$\text{Var}(X) = E[X^2] - (E[X])^2$.

For two random variables $X$ and $Y$, the covariance is defined as 
$\text{Cov}(X, Y) = E[(X-E[X])(Y-E[Y])]$.
It is easy to see that
$\text{Cov}(X, Y) = E[XY]-E[X]E[Y]$.
Then we define the correlation of $X$ and $Y$ as
\begin{equation}\label{classicalcorr}
\text{Corr}(X, Y) = \frac{\text{Cov}(X,Y)}{\sqrt{\text{Var}(X)\text{Var}(Y)}}
\end{equation}
It is also referred to the Pearson product-moment correlation coefficient or the bivariate correlation, as a measure for the linear correlation between $X$ and $Y$. 

\subsection{Correlations in statistical mechanics}

In statistical mechanics~\cite{sethna2006statistical}, the equilibrium correlation function for two random variables $S_1$ at position $\mathbf{x}$ and time $t$ and $S_2$ at position $\mathbf{x}+\mathbf{r}$ and time $t+\tau$ is defined as
\begin{equation}
C(\mathbf{r}, \tau) = \langle S_1(\mathbf{x}, t) S_2(\mathbf{x}+\mathbf{r}, t+\tau) \rangle - \langle S_1(\mathbf{x}, t) \rangle  \langle S_2(\mathbf{x}+\mathbf{r}, t+\tau) \rangle,
\end{equation}
where $\langle O \rangle$ is the thermal average of the random variable $O$; it is usually averaged over the whole phase space of the system. That is, 
\begin{equation}
\langle O \rangle = \frac{\int O e^{-\beta H(q_1, \dots, q_m, p_1, \dots, p_n)}\textrm{d}\tau}{\int e^{-\beta H(q_1, \dots, q_m, p_1, \dots, p_n)}\textrm{d}\tau},
\end{equation}
where $\beta = 1/k_BT$, $k_B$ is Boltzmann constant and $T$ is the temperature, $H$ is the Hamiltonian of the classical system in terms of coordinates $q_i$ and their conjugate generalised momenta $p_i$, and $\textrm{d}\tau$ is the volume element of the classical phase space. 

\section{Pseudo-density matrix formalism}
We firstly introduce the pseudo-density matrix approach ~\cite{fitzsimons2015quantum, zhao2018geometry, pisarczyk2019causal, zhang2020different} for defining quantum states over both space and time by treating quantum correlations in space and time equally.  
We review the definition of pseudo-density matrices for finite dimensions, continuous variables, and general measurement processes, and present their properties. 

\subsection{Finite dimensions: definition and properties}

The pseudo-density matrix formalism is originally proposed as a finite-dimensional quantum-mechanical formalism which aims to treat space and time on an equal footing~\cite{fitzsimons2015quantum}. 
In general, this formulation defines an event via making a measurement in space-time and is built upon correlations from measurement results; thus, it treats temporal correlations just as spatial correlations from observation of measurements and unifies spatio-temporal correlations in a single framework. 
As a price to pay, the spacetime states represented by pseudo-density matrices may not be positive semi-definite. 

An $n$-qubit density matrix can be expanded by Pauli operators $\sigma_i$ in terms of Pauli correlations which are the expectation values of these Pauli operators. 
In spacetime, instead of considering $n$ qubits, let us pick up $n$ events; for each event a single-qubit Pauli operator is measured. The pseudo-density matrix is then defined as 
\begin{equation}\label{pdm}
\hat{R} \equiv \frac{1}{2^n} \sum_{i_1=0}^{3}...\sum_{i_n=0}^{3} \langle \{\sigma_{i_j}\}_{j=1}^{n} \rangle \bigotimes_{j=1}^n \sigma_{i_j},
\end{equation}
where $\langle \{\sigma_{i_j}\}_{j=1}^{n} \rangle$ is the expectation value of the product of these measurement results for a particular choice of events with measurement operators $\{\sigma_{i_j}\}_{j=1}^{n}$.
Similar to a density matrix, a pseudo-density matrix is Hermitian and unit-trace; but it is not positive semi-definite as we mentioned before. 
If the measurements are space-like separated or local systems evolve independently, the pseudo-density matrix will reduce to a standard density matrix. 
Otherwise, for example if measurements are made in time, the pseudo-density matrix may have negative eigenvalues. 

Consider the bipartite case in time; that is, a single qubit $\rho$ at $t_A$ evolves to time $t_B$ under a quantum channel $\mathcal{E}: \rho \rightarrow \mathcal{E}(\rho)$. The pseudo-density matrix is given as 
\begin{align}
R & = (\mathcal{I} \times \mathcal{E})\left(\frac{1}{2}[\rho_A \otimes \frac{\mathbbm{1}_B}{2} S + S \rho_A \otimes \frac{\mathbbm{1}_B}{2}]\right)	 \nonumber\\ 
& = \frac{1}{2} \left(\rho_A \otimes \frac{\mathbbm{1}_B}{2}E_{AB} + E_{AB} \rho_A \otimes \frac{\mathbbm{1}_B}{2}\right)
\end{align}
where $E_{AB} =  (\mathcal{I} \times \mathcal{E})(\sum_{ij} \ket{i}\bra{j}\otimes\ket{j}\bra{i})$ is the Choi isomorphism of the quantum channel $\mathcal{E}$~\cite{horsman2017can}.

\subsection{Generalisation of pseudo-density matrix formalism}

The pseudo-density matrix formalism in continuous variables is given in various forms in Ref~\cite{zhang2020different}, including the Gaussian case, spacetime Wigner functions and corresponding spacetime density matrices, and for position measurements and weak measurements.

Gaussian states are fully characterised by the first two statistical moments of the quantum states, the mean value and the covariance matrix. 
The mean value $\bm{d}$, is defined as the expectation value of the $N$-mode quadrature field operators $\{\hat{q}_k, \hat{p}_k\}_{k=1}^N$ arranged in $\bm{\hat{x}} = (\hat{q}_1, \hat{p}_1, \cdots, \hat{q}_N, \hat{p}_N)^T$, that is, 
\begin{equation}
d_j= \langle \hat{x}_j \rangle _{\rho} \equiv \Tr (\hat{x}_j \hat{\rho}),
\end{equation}
for the Gaussian state $\hat{\rho}$.
The elements in the covariance matrix $\bm{\sigma}$ are defined as 
\begin{equation}
\sigma_{ij} = \langle \hat{x}_i \hat{x}_j + \hat{x}_j \hat{x}_i \rangle_{\rho} - 2 \langle \hat{x}_i \rangle_{\rho}\langle \hat{x}_j \rangle_{\rho}.
\end{equation}
A Gaussian \emph{spacetime state} is defined in Ref.~\cite{zhang2020different} via measurement statistics as being 
(i) a vector $\bm{d}$ of 2N expectation values of the $N$-mode quadrature field operators $\{\hat{q}_k, \hat{p}_k\}_{k=1}^N$ arranged in $\bm{\hat{x}} = (\hat{q}_1, \hat{p}_1, \cdots, \hat{q}_N, \hat{p}_N)^T$, with j-th entry
\begin{equation}\label{defmv}
d_j= \langle \hat{x}_j \rangle _{\rho}= \Tr (\hat{x}_j \rho).
\end{equation} and (ii) a covariance matrix $\bm{\sigma}$ with entries as
\begin{equation}\label{defcm}
\sigma_{ij} = 2 \langle\{\hat{x}_i, \hat{x}_j\}\rangle_{\rho} - 2 \langle \hat{x}_i \rangle_{\rho} \langle \hat{x}_j \rangle_{\rho}
\end{equation}
where $\langle \{\hat{x}_i, \hat{x}_j\}\rangle_{\rho}$ is the expectation value for the product of measurement results; specifically $\{\hat{x}_i, \hat{x}_j\} = \frac{1}{2} (\hat{x}_i \hat{x}_j + \hat{x}_j \hat{x}_i )$ for measurements at the same time. 
For general continuous variables and general measurement processes, see Ref.~\cite{zhang2020different}.

\section{Indefinite causal structures}
The concept of indefinite causal structures was proposed as probabilistic theories with non-fixed causal structures as a possible approach to quantum gravity~\cite{hardy2007towards, hardy2009quantum}. There are different indefinite causal order approaches: quantum combs~\cite{chiribella2008quantum, chiribella2009theoretical}, operator tensors~\cite{hardy2012operator, hardy2018construction}, process matrices~\cite{oreshkov2012quantum, araujo2015witnessing}, process tensors~\cite{milz2017introduction, pollock2018non}, and super-density operators~\cite{cotler2018superdensity, cotler2019quantum}. 
Also, Several of the approaches are closely related~\cite{costa2018unifying}, for example, quantum channels with memories~\cite{kretschmann2005quantum}, general quantum strategies~\cite{gutoski2007toward}, multiple-time states~\cite{aharonov1964time, aharonov2009multiple, silva2017connecting}, general boundary formalism~\cite{oeckl2003general}, and quantum causal models~\cite{costa2016quantum, allen2017quantum}. General quantum strategies can be taken as a game theory representation; multiple-time states are a particular subclass of process matrices; quantum causal models just use the process matrix formalism.
Since there are clear maps among quantum combs, operator tensors, process tensors, and process matrices, we just take the process matrix formalism in order to learn from causality inequalities, quantum switch and post-selection.
We will investigate its relation with the pseudo-density matrix and show what lessons we shall learn for pseudo-density matrices. 

\subsection{Preliminaries for process matrix formalism}

The process matrix formalism is originally proposed in Ref.~\cite{oreshkov2012quantum} as one of the indefinite causal structures assuming local quantum mechanics and well-defined probabilities. The process matrix was defined to take completely positive(CP) maps to linear probabilities. 
It is redefined in Ref.~\cite{araujo2017purification} in a more general way as high order transformations, where the definition is extended to take CP maps to other CP maps. 
Here we follow as Ref.~\cite{araujo2017purification}. We define bipartite processes first; the multipartite case is obtained directly or from Ref.~\cite{araujo2015witnessing}.

For the bipartite case, consider a global past $P$ and a global future $F$. Quantum states in the past are transformed to quantum states in the future through a causally indefinite structure. 
 A process is defined as a linear transformation take two CPTP maps $\mathcal{A}: A_I \otimes A_I' \rightarrow A_O \otimes A_O'$ and $\mathcal{B}: B_I \otimes B_I' \rightarrow B_O \otimes B_O'$ to a CPTP map $\mathcal{G}_{\mathcal{A},\mathcal{B}}: A_I' \otimes B_I' \otimes P \rightarrow A'_O \otimes B_O' \otimes F$ without acting on $A_I'$, $A_O'$, $B_I'$, $B_O'$. Specifically, it is a transformation that act on $P \otimes A_I \otimes A_O \otimes B_I \otimes B_O \otimes F$.
 
 We introduce the Choi-Jamio\l{}kowski isomorphism~\cite{jamiolkowski1972linear, choi1975completely} to represent the process in the matrix formalism. 
 Recall that for a completely positive map $\mathcal{M}^A: A_I \rightarrow A_O$, its corresponding Choi-Jamio\l{}kowski matrix is given as $\mathfrak{C}(\mathcal{M}) \equiv [\mathcal{I} \otimes \mathcal{M}^A (|\mathbbm{1} \rrangle \llangle \mathbbm{1} |)]\in A_I \otimes A_O$ with $\mathcal{I}$ as the identity map and $|\mathbbm{1} \rrangle = |\mathbbm{1} \rrangle^{A_I A_I} \equiv \sum_j \ket{j}^{A_I} \otimes \ket{j}^{A_I} \in \mathcal{H}^{A_I} \otimes \mathcal{H}^{A_I}$ is the non-normalised maximally entangled state. The inverse is given as $\mathcal{M}(\rho^{A_I}) = \Tr[ (\rho^{A_I}\otimes \mathbbm{1}^{A_O}) M^{A_IA_O}]$ where $\mathbbm{1}^{A_O}$ is the identity matrix on $\mathcal{H}^{A_O}$.

Then $A = \mathfrak{C}(\mathcal{A})$, $B = \mathfrak{C}(\mathcal{B})$, and $G_{A, B} = \mathfrak{C}(\mathcal{G_{A,B}})$ are the corresponding CJ representations. We have
\begin{equation}
G_{A, B} = 	\Tr_{A_IA_OB_IB_O}[W^{T_{A_IA_OB_IB_O}}(A\otimes B)]
\end{equation}
where the process matrix is defined as $W \in P \otimes A_I \otimes A_O \otimes B_I \otimes B_O \otimes F$, $T_{A_IA_OB_IB_O}$ is the partial transposition on the subsystems $A_I$, $A_O$, $B_I$, $B_O$, and we leave identity matrices on the rest subsystems implicit. 
Note that we require that $G_{A, B}$ is a CPTP map for any CPTP maps $A$, $B$. This condition is equivalent to the followings:
\begin{align}
W \geq 0, \\
\Tr W = d_{A_O}d_{B_O}d_P, \\
W = L_V(W),
\end{align}
where $L_V$ is defined as a projector 
\begin{align}
L_V(W) = W & - _FW + _{A_OF}W	+ _{B_OF}W \nonumber \\
&- _{A_OB_OF}W - _{A_IA_OF}W + _{A_IA_OB_OF}W \nonumber \\
& - _{B_IB_OF}W + _{A_IA_OB_OF}W - _{A_IA_OB_IB_OF}W\nonumber \\
&  + _{PA_IA_OB_IB_OF}W.
\end{align}

Terms that can exist in a process matrix include states, channels, channels with memory; nevertheless, postselection, local loops, channels with local loops and global loops are not allowed~\cite{oreshkov2012quantum}.
A bipartite process matrix can be fully characterised in the Hilbert-Schmidt basis~\cite{oreshkov2012quantum}. 
Define the signalling directions $\preceq$ and $\npreceq$ as follows: $A \preceq B$ means $A$ is in the causal past of $B$, $A \npreceq B$ means it is not; similar for  $\succeq$ and $\nsucceq$. 
Any valid bipartite process matrix $W^{A_IA_OB_IB_O}$ can be given in the Hilbert-Schmidt basis as
\begin{equation}
W^{A_IA_OB_IB_O} = \frac{1}{d_{A_I}d_{B_I}} (\mathbbm{1} + \sigma_{A \preceq B} + \sigma_{A \succeq B} + \sigma_{A \npreceq \nsucceq B})
\end{equation}
where 
the matrices $\sigma_{A \preceq B}$, $\sigma_{A \succeq B}$, and $\sigma_{A \npreceq \nsucceq B}$ are defined by 
\begin{align}
\sigma_{A \preceq B} & \equiv \sum_{ij>0} c_{ij}\sigma_i^{A_O}\sigma_j^{B_I} + \sum_{ijk>0} d_{ijk}\sigma_i^{A_I}\sigma_j^{A_O}\sigma_k^{B_I}\\
\sigma_{A \succeq B} & \equiv \sum_{ij>0} e_{ij}\sigma_i^{A_I}\sigma_j^{B_O} + \sum_{ijk>0} f_{ijk}\sigma_i^{A_I}\sigma_j^{B_I}\sigma_k^{B_O}\\
\sigma_{A \npreceq \nsucceq B} & \equiv \sum_{i>0} g_i\sigma_i^{A_I} + \sum_{i>0} h_i\sigma_i^{B_I} + \sum_{ij>0} l_{ij}\sigma_i^{A_I}\sigma_j^{B_I} \\
\end{align}
Here $c_{ij}, d_{ijk}, e_{ij}, f_{ijk}, g_i, h_i,  l_{ij} \in \mathbb{R}$. 
That is, a bipartite process matrix of the system $AB$ is a combination of an identity matrix, the matrices where $A$ signals to $B$, where $B$ signals to $A$, and where $A$ and $B$ are causally separated. It is thus a linear combination of three possible causal structures. 





\subsection{Correlation analysis}

Now we analyse the relation between a process matrix and a pseudo-density matrix in finite dimensions. The basic elements in a process matrix are different laboratories, and the basic elements in a pseudo-density matrix are different events. We map a process matrix to a pseudo-density matrix in a way that each lab corresponds to each event. 

A process matrix with a single-qubit Pauli measurement taken at each laboratory is mapped to a finite-dimensional pseudo-density matrix. 
Compare them in the bipartite case as an illustration. 
In the simplest temporal case, a maximally mixed qubit evolves under the identity evolution between two times. 
The process matrix for this scenario is given as 
\begin{equation}
W = \frac{\mathbbm{1}^{A_I}}{2} \otimes [[\mathbbm{1}]]^{A_OB_I}, 
\end{equation}
where $[[\mathbbm{1}]]^{XY} = \sum_{ij}\ket{i}\bra{j}^X \otimes \ket{i}\bra{j}^Y = \frac{1}{2}(\mathbbm{1} \otimes \mathbbm{1} + X \otimes X - Y \otimes Y + Z\otimes Z)$. 
At the same time, the corresponding pseudo-density matrix is
\begin{equation}
R = \frac{1}{4} (I\otimes I + X \otimes X + Y \otimes Y + Z\otimes Z)= \frac{1}{2}[[\mathbbm{1}]]^{PT}= \frac{1}{2} S, 
\end{equation}
where the swap operator $S = \frac{1}{2}(\mathbbm{1}\otimes \mathbbm{1}  + X \otimes X + Y \otimes Y + Z\otimes Z) = [[\mathbbm{1}]]^{PT}$, here $PT$ is the partial transpose. 
For an arbitrary state $\rho$ evolving under the unitary evolution $U$, the process matrix is given as 
\begin{equation}
W = \rho^{A_I} \otimes [[U]]^{A_OB_I}, 
\end{equation}
where $[[U]] = (\mathbbm{1} \otimes U) [[\mathbbm{1}]] (\mathbbm{1} \otimes U^{\dag})$. The pseudo-density matrix is given from Ref.~\cite{zhao2018geometry} as
\begin{align}
R & = \frac{1}{2} (\mathbbm{1} \otimes U) (\rho^A \otimes \frac{\mathbbm{1}^B}{ 2} S + S\rho^A \otimes \frac{\mathbbm{1}^B}{2}) (\mathbbm{1} \otimes U^{\dag}) \nonumber \\
& = \frac{1}{2}(\rho^A \otimes \frac{\mathbbm{1}^B}{2} [[U]]^{PT} + [[U]]^{PT} \rho^A \otimes \frac{\mathbbm{1}^B}{2}),
\end{align}
where the partial transpose is taken on the subsystem $A$. 
Now we compare the correlations in the two formalisms and check whether they hold the same information.

The single-qubit Pauli measurement $\sigma_{i}$ for each event in the pseudo-density matrix has the Choi-Jamio\l{}kowski representation as 
\begin{equation}
\Sigma^{A_IA_O}_{i} = P^{+ A_I}_i \otimes P^{+ A_O}_i - P^{- A_I}_i\otimes P^{- A_O}_i  
\end{equation} 
where $P^{\pm}_i = \frac{1}{2}(\mathbbm{1} \pm \sigma_i)$; that is, to make a measurement $P^{\alpha}_i (\alpha = \pm 1)$ to the input state and project the corresponding eigenstate to the output system. 
It is equivalent to
\begin{equation}
\Sigma^{A_IA_O}_{i} = \frac{1}{2} (\mathbbm{1}^{A_I} \otimes \sigma_i^{A_O} + \sigma_i^{A_I} \otimes \mathbbm{1}^{A_O}).
\end{equation} 
In the example of a single qubit $\rho$ evolving under $U$, the correlations from the process matrix are given by
\begin{align}
p( \Sigma^{A_IA_O}_{i}, \Sigma^{B_IB_O}_{j}) & = \Tr [(\Sigma^{A_IA_O}_{i} \otimes \Sigma^{B_IB_O}_{j} )W] \nonumber \\
& = \frac{1}{2} \Tr [\sigma_j U \sigma_i U^{\dag}];
\end{align}
while the correlations from the pseudo-density matrix are given as 
\begin{align}
\langle \{ \sigma_i, \sigma_j \} \rangle &= \frac{1}{2}\left( \Tr[\sigma_j U \sigma_i \rho U^{\dag} ] + \Tr[\sigma_j U \rho \sigma_i U^{\dag} ] \right)\nonumber \\
&= \frac{1}{2} \Tr [\sigma_j U \sigma_i U^{\dag}].
\end{align}
The last equality holds as a single-qubit $\rho$ is decomposed into $\rho = \frac{\mathbbm{1}}{2} + \sum_{k=1, 2, 3} c_k \sigma_k$. 
The allowed spatio-temporal correlations given by the two formalisms are the same; thus, pseudo-density matrices and process matrices are equivalent in terms of encoded correlations. 
In a general case of bipartite systems on $AB$, this equivalence holds for $A \preceq B$, $A \succeq B$, $A \npreceq \nsucceq B$ and thus their superpositions for arbitrary process matrices. 
The only condition is that $A$ and $B$ make Pauli measurements in their local laboratories. 
Therefore, a process matrix where a single-qubit Pauli measurement is made at each laboratory corresponds to a finite-dimensional pseudo-density matrix since the correlations are equal. 

For generalised measurements, for example, arbitrary POVMs, a process matrix is fully mapped to the corresponding generalised pseudo-density matrix; thus, a process matrix can be always mapped to a generalised pseudo-density matrix in principle. 
 The process matrix and the corresponding generalised pseudo-density matrix just take the same measurement process in each laboratory or at each event. 
The analysis for correlations is similar. 

For a given set of measurements, a process matrix where the measurement is made in each laboratory hold the same correlations as a generalised pseudo-density matrix with the measurement made at each event. 
Thus, a universal mapping from a process matrix to a pseudo-density matrix for general measurements is established. 

However, a pseudo-density matrix in finite dimensions is not necessarily mapped back to a valid process matrix. As mentioned before, a valid process matrix excludes the possibilities for post-selection, local loops, channels with local loops and global loops. Pseudo-density matrices are defined operationally in terms of measurement correlations and may allow these possibilities. 
We will come back to this point in the discussion for post-selection and out-of-time-order correlation functions.

\subsection{Causal inequalities}
In the subsection, we introduce the causal polytope formed by the set of correlations with a definite causal order. Its facets are defined as causal inequalities.~\cite{branciard2015simplest} We show that the characterisation of bipartite correlations is consistent with previous analysis in the pseudo-density matrix formalism. We show that causal inequalities can be violated in both of process matrix formalism and pseudo-density matrix formalism. 

Here we follow from Ref.~\cite{branciard2015simplest}. Recall that we denote Alice in the causal past of Bob as $A \preceq B$. Now for simplicity, we do not consider relativistic causality but normal Newton causality. We denote $A \prec B$ for events in Alice's system precedes those in Bob's system. Then Bob cannot signal to Alice, and the correlations satisfy that 
\begin{equation}\label{nosignalingtoAlice}
\forall x, y, y', a, \quad p^{A \prec B}(a|x,y) = p^{A \prec B}(a|x, y'),	
\end{equation}
where $p^{A \prec B}(a|x,y^{(')}) = \sum_b p^{A \prec B}(a,b|x,y^{(')})$.
Similarly, for $B \prec A$, Alice cannot signal to Bob that
\begin{equation}\label{nosignalingtoAlice}
\forall x, x', y, b, \quad p^{A \prec B}(b|x,y) = p^{A \prec B}(b|x', y),	
\end{equation}
where $p^{A \prec B}(b|x^{(')}, y) = \sum_a p^{A \prec B}(a,b|x^{(')}, y)$.

Correlations of the order $A \prec B$ satisfy non-negativity, normalisation, and the no-signaling-to-Alice condition:
\begin{align}
p^{A \prec B}(a,b|x,y) \geq &0, \quad\forall x,y,a,b;\\
\sum_{a,b}p^{A \prec B}(a,b|x,y) = &1, \quad\quad\forall x,y;\\
p^{A \prec B}(a|x,y) = p^{A \prec B}(a|&x, y'), \quad\forall x,y,y',a.
\end{align}
The set of correlations $p^{A \prec B}$ forms a convex polytope. 
Similarly for the set of correlations $p^{B \prec A}$.
The correlations are defined as causal if it is compatible with $A \prec B$ with probability $q$ and $B \prec A$ with probability $1-q$, that is for $q \in [0,1]$, 
\begin{equation}
p(a,b|x,y) = q p^{A \prec B}(a,b|x,y) + (1-q) p^{B \prec A}(a,b|x,y),
\end{equation}
where $p^{A \prec B}$ and $p^{B \prec A}$ are non-negative and normalised to 1.
The set of causal correlation is the convex hull of  $p^{A \prec B}$ and $p^{B \prec A}$ and constitutes the causal polytope.

Suppose that Alice and Bob's inputs have $m_A$ and $m_B$ possible values, their outputs have $k_A$ and $k_B$ values respectively. 
The polytope of $p^{A \prec B}$ has $k_A^{m_A}k_B^{m_Am_B}$ vertices, of dimension $m_Am_B(k_Ak_B-1) - m_A(m_B-1)(k_A-1)$.
The polytope of $p^{B \prec A}$ has $k_A^{m_Am_B}k_B^{m_B}$ vertices, of dimension $m_Am_B(k_Ak_B-1) - (m_A-1)m_B(k_A-1)$.
The causal polytope has $k_A^{m_A}k_B^{m_Am_B} + k_A^{m_Am_B}k_B^{m_B} - k_A^{m_A}k_B^{m_B}$ vertices, of dimension $m_Am_B(k_Ak_B-1)$.
Consider the bipartite correlations. 
For example, a qubit evolves between two times $t_A$ and $t_B$. We make a Pauli measurement at each time to record correlations. Given an initial state of the qubit, we have $m_A = m_B = 1$, $k_A = k_B = 2$. The polytope of $p^{A \prec B}$ has 4 vertices in 3 dimensions. The same as $p^{B \prec A}$ and the causal polytope. This result is consistent with the characterisation by pseudo-density matrix formalism in Ref.~\cite{zhao2018geometry}. 

Now we characterise the causal polytope with $m_A = m_B = k_A = k_B = 2$. 
It has 112 vertices and 48 facets. 16 of the facets are trivial, which imply the non-negativity of the correlations $p(a,b|x,y)\geq 0$.
If we relabel the inputs and outputs of the systems, the rest facets are divided into two groups, each with 16 facets:
\begin{equation}\label{gyni}
\frac{1}{4} \sum_{x,y,a,b} \delta_{a,y}\delta_{b,x}p(a,b|x,y) \leq \frac{1}{2},
\end{equation}
and 
\begin{equation}\label{lgyni}
\frac{1}{4} \sum_{x,y,a,b} \delta_{x(a\oplus y), 0}\delta_{y(b\oplus x), 0}p(a,b|x,y) \leq \frac{3}{4},
\end{equation}
where $\sigma_{i,j}$ is the Kronecker delta function and $\oplus$ is the addition modulo 2.
They are interpreted into bipartite "guess your neighbour's input" (GYNI) games and "lazy GYNI" (LGYNI) games~\cite{branciard2015simplest}.

Then we show the violation of causal inequalities via process matrix formalism and pseudo-density matrix formalism. 
In the process matrix formalism, we take the global past $P$, the global future $F$, Alice's ancilla systems $A_I'$, $A_O'$ and Bob's ancilla systems $B_I'$, $B_O'$ trivial. 
Then the process matrix correlations are given as 
\begin{equation}
p(a, b|x,y) = \Tr[ W^{T_{A_IA_OB_IB_O}} A_{a|x} \otimes B_{b|y}]	.
\end{equation}
Consider the process matrix 
\begin{equation}
W = \frac{1}{4}\left[ \mathbbm{1}^{\otimes 4} + \frac{Z^{A_I}Z^{A_O}Z^{B_I}\mathbbm{1}^{B_O} + Z^{A_I}\mathbbm{1}^{A_O}X^{B_I}X^{B_O}}{\sqrt{2}} \right].
\end{equation}
We choose the operations as (here slightly different from Ref.~\cite{branciard2015simplest}):
\begin{align}
A_{0|0} &= B_{0|0} = 0,\\
A_{1|0} &= B_{1|0} = (\ket{00}+\ket{11})(\bra{00}+\bra{11}), \\
A_{0|1} &= B_{0|1} = \frac{1}{2} \ket{0}\bra{0} \otimes \ket{0}\bra{0}+ \frac{1}{2}\ket{0}\bra{0} \otimes \ket{1}\bra{1}, \\
A_{1|1} &= B_{1|1} = \frac{1}{2}\ket{1}\bra{1} \otimes \ket{0}\bra{0}+ \frac{1}{2}\ket{1}\bra{1} \otimes \ket{1}\bra{1}.
\end{align}
Then 
\begin{align}
p_{GYNI} & = \frac{5}{16}(1+ \frac{1}{\sqrt{2}}) \approx 0.5335 > \frac{1}{2},\\
p_{LGYNI} & = \frac{5}{16}(1+ \frac{1}{\sqrt{2}}) + \frac{1}{4} \approx 0.7835 > \frac{3}{4}.
\end{align}

For a pseudo-density matrix, we consider a similar strategy.
Alice has two systems $X$ and $A$, where $X$ is the ancillary system prepare with $\ket{x}\bra{x}$. Bob has two systems $Y$ and $B$, where $Y$ is the ancillary system prepare with $\ket{y}\bra{y}$.
Given a pseudo-density matrix 
\begin{align}
R = \frac{1}{4}&[ \ket{x}\bra{x}^X \otimes \mathbbm{1}^A \otimes \ket{y}\bra{y}^Y \otimes \mathbbm{1}^B \nonumber \\
& + \frac{Z^{X}Z^{A}Z^{Y}\mathbbm{1}^{B} + Z^{X}\mathbbm{1}^{A}X^{Y}X^{B}}{\sqrt{2}}],
\end{align}
we choose the operations as before and gain again
\begin{align}
p_{GYNI} & = \frac{5}{16}(1+ \frac{1}{\sqrt{2}}) \approx 0.5335 > \frac{1}{2},\\
p_{LGYNI} & = \frac{5}{16}(1+ \frac{1}{\sqrt{2}}) + \frac{1}{4} \approx 0.7835 > \frac{3}{4}.
\end{align}

Again the causal inequalities are violated. 
This example also highlights another relationship for the mapping between a process matrix and a pseudo-density matrix. Instead of an input system and an output system in a process matrix, the corresponding pseudo-density matrix has an additional ancillary system for each event. 

A process matrix which makes a measurement and reprepares the state in one laboratory describes the same probabilities as a pseudo-density matrix with ancillary systems which makes a measurement and reprepares the state at each event. 
Thus, another mapping from a process matrix to a pseudo-density matrix is established by introducing ancillary systems.

\subsection{Postselection}

Post-selection is conditioning on the occurrence of certain event in probability theory, or conditioning upon certain measurement outcome in quantum mechanics. 
It allows a quantum computer to choose the outcomes of certain measurements and increases its computational power significantly.
In this subsection, we take the view from post-selection and show that a particular subset of post-selected two-time states correspond to process matrices in indefinite causal order. Post-selected closed timelike curves are presented as a special case.

\subsubsection{Two-time quantum states}

In this subsubsection, we review the two-time quantum states approach~\cite{silva2017connecting} which fixes independent initial states and final states at two times. The two-time quantum state takes its operational meaning from post-selection. Consider that Alice prepares a state $\ket{\psi}$ at the initial time $t_1$. Between the initial time $t_1$ and the final time $t_2$, she performs arbitrary operations in her lab. Then she measures an observable $O$ at the final time $t_2$.  The observable $O$ has a non-degenerate eigenstate $\ket{\phi}$. Taking $\ket{\phi}$ as the final state, Alice discards the experiment if the measurement of $O$ does not give the eigenvalue corresponding to the eigenstate $\ket{\phi}$. 

Consider that Alice makes a measurement by the set of Kraus operators $\{\hat{E}_a = \sum_{k,l} \beta_{a, kl}\ket{k}\bra{l}\}$ between $t_1$ and $t_2$. Note that $\{\hat{E}_a\}$ are normalised as $\sum_a \hat{E}_a^{\dag}\hat{E}_a = \mathbbm{1}$. The probability for Alice to gain the outcome $a$ under the pre- and post-selection is given as
\begin{equation}
p(a) = \frac{|\bra{\phi}\hat{E}_a\ket{\psi}|^2}{\sum_{a'} |\bra{\phi}\hat{E}_{a'}\ket{\psi}|^2}. 
\end{equation}
Now define the two-time state and the two-time version of Kraus operator as
\begin{align}
\Phi = _{\mathcal{A}_2}\bra{\phi} \otimes \ket{\psi}^{\mathcal{A}_1} & \in \mathcal{H}_{\mathcal{A}_2} \otimes \mathcal{H}^{\mathcal{A}_1}, \nonumber \\ 
E_a = \sum_{kl} \beta_{a, kl} \ket{k}^{\mathcal{A}_2} \otimes _{\mathcal{A}_1}\bra{l} & \in \mathcal{H}^{\mathcal{A}_2} \otimes \mathcal{H}_{\mathcal{A}_1},
\end{align}
where the two-time version of Kraus operator is denoted by $E_a$ without the hat. 
An arbitrary pure two-time state takes the form
\begin{equation}
\Phi = \sum \alpha_{ij} \ _{\mathcal{A}_2}\bra{i} \otimes \ket{j}^{\mathcal{A}_1} \in \mathcal{H}_{\mathcal{A}_2} \otimes \mathcal{H}^{\mathcal{A}_1}.
\end{equation}
Then the probability to obtain $a$ as the outcome is given as 
\begin{equation}
p(a) = \frac{|\Phi \cdot E_a |^2}{\sum_{a'} |\Phi \cdot E_{a'}|^2}. 
\end{equation}

A two-time density operator $\eta$ is given as 
\begin{equation}
\eta = \sum_r p_r \Phi_r \otimes \Phi_r^{\dag} \in \mathcal{H}_{\mathcal{A}_2} \otimes \mathcal{H}^{\mathcal{A}_1} \otimes \mathcal{H}_{\mathcal{A}_1^{\dag}} \otimes \mathcal{H}^{\mathcal{A}_2^{\dag}}.
\end{equation}
Consider a coarse-grained measurement 
\begin{equation}
J_a  = \sum_{\mu} E_a^{\mu} \otimes E_a^{\mu \dag} \in \mathcal{H}^{\mathcal{A}_2} \otimes \mathcal{H}_{\mathcal{A}_1}  \otimes \mathcal{H}^{\mathcal{A}_1^{\dag}} \otimes \mathcal{H}_{\mathcal{A}_2^{\dag}} 
\end{equation}
where the outcome $a$ corresponds to a set of Kraus operators $\{\hat{E}_a^{\mu}\}$. 
Then the probability to obtain $a$ as the outcome is given as 
\begin{equation}
p(a) = \frac{\eta \cdot J_a}{\sum_{a'} \eta \cdot J_{a'}}. 
\end{equation}

\subsubsection{Connection between process matrix and pseudo-density matrix under post-selection}


Now consider post-selection applied to ordinary quantum theory. 
It is known that a particular subset of post-selected two-time states in quantum mechanics give the form of process matrices within indefinite causal structures~\cite{silva2017connecting}. 
Here we first give a simple explanation for this fact and further analyse the relation between a process matrix and a pseudo-density matrix from the view of post-selection. 

For an arbitrary bipartite process matrix $W \in \mathcal{H}^{A_I} \otimes \mathcal{H}^{A_O} \otimes \mathcal{H}^{B_I} \otimes \mathcal{H}^{B_O} $, we can expand it in some basis: 
\begin{equation}
W^{A_IA_OB_IB_O} = \sum_{ijkl, pqrs} w_{ijkl, pqrs} \ket{ijkl}\bra{pqrs}.
\end{equation}
For the elements in each Hilbert space, we map them to the corresponding parts in a bipartite two-time state. For example, we map the input Hilbert space of Alice to the bra and ket space of Alice at time $t_1$, and similarly for the output Hilbert space for $t_2$. That is, 
\begin{align}
\ket{i}\bra{p} & \in \mathcal{L}(\mathcal{H}^{A_I}) \rightarrow \bra{p} \otimes \ket{i} \in \mathcal{H}_{A_1^{\dag}} \otimes \mathcal{H}^{A_1}\\
\ket{j}\bra{q} & \in \mathcal{L}(\mathcal{H}^{A_O})\rightarrow \bra{q} \otimes \ket{j} \in \mathcal{H}_{A_2} \otimes \mathcal{H}^{A_2^{\dag}}
\end{align}
Thus, a two-time state $\eta_{W^{A_1A_2}} \in \mathcal{H}_{A_2} \otimes \mathcal{H}^{A_1} \otimes \mathcal{H}^{A_2^{\dag}} \otimes \mathcal{H}_{A_1^{\dag}}$ is equivalent to a process matrix for a single laboratory $W^{A_IA_O}$.

The connection with pre- and post-selection suggests one more interesting relationship between a process matrix and a pseudo-density matrix. 
For a process matrix, if we consider the input and output Hilbert spaces at two times, we can map it to a two-time state. That is, we connect a process matrix with single laboratory to a two-time state. 
A pseudo-density matrix needs two Hilbert spaces to represent two times. For a two-time state $\eta_{12}$, the corresponding pseudo-density matrix $R_{12}$ has the same marginal single-time states, i.e., $\Tr_1 \eta_{12} = \Tr_1 R_{12}$ and $\Tr_2 \eta_{12} = \Tr_2 R_{12}$.
Then we find a map between a process matrix for a single event and a pseudo-density matrix for two events. 
Note that in the previous subsections, we have mapped a process matrix for two events to one pseudo-density matrix with half Hilbert space for two events, and mapped a process matrix for two events to a pseudo-density matrix with two Hilbert spaces at each of two events. 
This suggests that the relationship between a process matrix and a pseudo-density matrix is non-trivial with a few possible mappings. 


One question arising naturally here concerns the pseudo-density matrices with post-selection. 
The definitions for finite-dimensional and Gaussian pseudo-density matrices guarantee that under the partial trace, the marginal states at any single time will give the state at that time. In particular, tracing out all other times in a pseudo-density matrix, we get the final state at the final time. On the one hand, we may think that pseudo-density matrix formulation is kind of time-symmetric. On the other hand, the final state is fixed by evolution; that implies that we cannot assign an arbitrary final state, making it difficult for the pseudo-density matrix to be fully time-symmetric. For other generalisation of pseudo-density matrices like position measurements and weak measurements, the property for fixed final states does not hold. 
Nevertheless, we may define a new type of pseudo-density matrices with post-selection. We assign the final measurement to be the projection to the final state and renormalise the probability. For example, a qubit in the initial state $\rho$ evolves under a CPTP map $\mathcal{E}: \rho \rightarrow \mathcal{E}(\rho)$ and then is projected on the state $\eta$. We may construct the correlations $\langle \{ \sigma_i , \sigma_j, \eta \} \rangle$ as 
\begin{equation}
\langle \{ \sigma_i , \sigma_j, \eta \} \rangle= \sum_{\alpha, \beta = \pm 1} \alpha \beta \Tr[\eta P^{\beta}_j \mathcal{E}(P^{\alpha}_i \rho P^{\alpha}_i )  P^{\beta}_j ] / p_{ij}(\eta),
\end{equation}
where $P^{\alpha}_i = \frac{1}{2}(\mathbbm{1} + \alpha \sigma_i)$ and $p_{ij}(\eta) = \sum_{\alpha, \beta = \pm 1} \Tr[\eta P^{\beta}_j \mathcal{E}(P^{\alpha}_i \rho P^{\alpha}_i )  P^{\beta}_j ]$. 
Then the pseudo-density matrix with post-selection is given as 
\begin{equation}
R = \frac{1}{4} \sum_{i, j = 0}^{3} \langle \{ \sigma_i , \sigma_j, \eta \} \rangle \sigma_i \otimes \sigma_j \otimes \eta.
\end{equation}
We further conclude the relation between a process matrix and a pseudo-density matrices with post-selection. 
A process matrix with postselection in a laboratory is operationally equivalent to a tripartite postselected pseudo-density matrix.

We briefly discuss post-selected closed timelike curves before we move on to a summary. 
Closed timelike curves (CTCs), after being pointed out by G\"{o}del to be allowed in general relativity~\cite{godel1949example}, have always been arising great interests. 
Deutsch~\cite{deutsch1991quantum} proposed a circuit method to study them and started an information theoretic point of view. Deustch's CTCs are shown to have many abnormal properties violated by ordinary quantum mechanics. For example, they are nonunitary, nonlinear, and allow quantum cloning~\cite{ahn2013quantum, brun2013quantum}. Several authors~\cite{bennett2005teleportation, svetlichny2011time, brun2012prefect, lloyd2011closed} later proposed a model for closed timelike curves based on post-selected teleportation.  
It is studied that process matrices correspond to a particular linear version of post-selected closed timelike curves~\cite{araujo2017quantum}. 
In pseudo-density matrices we can consider a system evolves in time and back; that is the case for calculating out-of-time-order correlation functions we will introduce later. 
For post-selected closed timelike curves, it is better to be illustrated by the pseudo-density matrices with post-selection. 

\subsection{Summary of the relation between pseudo-density matrix and indefinite causal structures}
In this section, we have introduced the relation between pseudo-density matrices and indefinite causal order. We argue that the pseudo-density matrix formalism belongs to indefinite causal structures. 
So far, all other indefinite causal structures to our knowledge use a tensor product of input and output Hilbert spaces, while a pseudo-density matrix only assumes a single Hilbert space. 
For a simple example of a qudit at two times, the dimension used in other indefinite causal structures is $d^4$ but for pseudo-density matrix is $2d^2$. Though other indefinite causal structures assume a much larger Hilbert space, pseudo-density matrix should not be taken as a subclass of any indefinite causal structures which already exist. There are certain non-trivial relation between pseudo-density matrices and other indefinite causal structures. 
As we can see from the previous subsections, 
a process matrix and the corresponding pseudo-density matrix allow the same correlations or probabilities in three different mappings.  
\begin{claim}
	It is possible to map a process matrix to a corresponding pseudo-density matrix under correlations in three different ways: one-lab to one-event direct map, one-lab to one-event with double Hilbert spaces map, and one-lab to two-event map. 
\end{claim}
One obvious difference between a process matrix and a pseudo-density matrix is that, for each laboratory, a process matrix measures and reprepares a state while a pseudo-density matrix usually only makes a measurement and the state evolves into its eigenstate for each eigenvalue with the corresponding probability. 
The correlations given by process matrices and pseudo-density matrices are also the same. 
Examples in post-selection and closed time curves suggest further similarities. 
In general, we can understand that the pseudo-density matrix is defined in an operational way which does not specify the causal order, thus belongs to indefinite causal structures. 
We borrow the lessons from process matrices here to investigate pseudo-density matrices further. Maybe it will be interesting to derive a unified indefinite causal structure which takes the advantage of all existing ones. 

Nevertheless, the ultimate goal of indefinite causal order towards quantum gravity is still far reaching. So far, all indefinite causal structures are linear superpositions of causal structures; will that be enough for quantising gravity?
It is generally believed among indefinite causal structure community that what is lacking in quantum gravity is the quantum uncertainty for dynamical causal structures suggested by general relativity.  
The usual causal order may be changed under this quantum uncertainty and there is certain possibility for a superposition of causal orders. 
It is an attractive idea; however, being criticised due to lack of evidence to justify the existence. 
One may argue that process matrices and quantum switch can describe part of the universe; however, such an approach to quantum gravity remains doubt. For example, indefinite causal structures restrict to linear superpositions of causal orders and can only describe linear post-selected closed timelike curves. Why is the ultimate theory of nature necessary to be linear?

\section{Consistent histories}

In this section we review on consistent histories and explore the relation between pseudo-density matrices and consistent histories. 

\subsection{Preliminaries for consistent histories}
Consistent histories, or decoherent histories, is an interpretation for quantum theory, proposed by Griffiths~\cite{griffiths1984consistent, griffiths2003consistent}, Gell-Mann and Hartle~\cite{gell2018quantum, gell1993classical}, and Omnes~\cite{omnes1990hilbert}. 
The main idea is that a history, understood as a sequence of events at successive times, has a consistent probability with other histories in a closed system. The probabilities assigned to histories satisfy the consistency condition to avoid the interference between different histories and that set of histories are called consistent histories~\cite{isham1995continuous, isham1998continuous, isham1995quantum, isham1994quantum, halliwell1994review, dowker1992quantum, dowker1996consistent}. 


Consider a set of projection operators $\{P_{\alpha}\}$ which are exhaustive and mutually exclusive:
\begin{equation}
\sum_{\alpha} P_{\alpha} = \mathbbm{1}, \qquad P_{\alpha}P_{\beta} = \delta_{\alpha\beta}P_{\beta},
\end{equation}
where the range of $\alpha$ may be finite, infinite or even continuous. 
For each $P_{\alpha}$ and a system in the state $\rho$, the event $\alpha$ is said to occur if $P_{\alpha}\rho P_{\alpha} = \rho$ and not to occur if $P_{\alpha}\rho P_{\alpha}=0$. 
The probability of the occurrence of the event $\alpha$ is given by
\begin{equation}
p(\alpha) = \Tr[P_{\alpha}\rho P_{\alpha}].
\end{equation}
A projection of the form $P_{\alpha} = \ket{\alpha}\bra{\alpha}$ ($\{\ket{\alpha}\}$ is complete) is called completely fine-grained, which corresponds to the precise measurement of a complete set of commuting observables. Otherwise, for imprecise measurements or incomplete sets, the projection operator is called coarse-grained. Generally it takes the form $\bar{P}_{\bar{\alpha}} = \sum_{\alpha \in \bar{\alpha}} P_{\alpha}$.

In the Heisenberg picture, the operators for the same observables $P$ at different times are related by
\begin{equation}
P(t) = \exp(iHt/\hbar) P(0) \exp(-iHt/\hbar),
\end{equation}
with $H$ as the Hamiltonian of the system.
Then the probability of the occurrence of the event $\alpha$ at time $t$ is
\begin{equation}
p(\alpha) = \Tr[P_{\alpha}(t) \rho P_{\alpha}(t)].
\end{equation}

Now we consider how to assign probabilities to histories, that is, to a sequence of events at successive times. 
Suppose that the system is in the state $\rho$ at the initial time $t_0$. Consider a set of histories $[\alpha] = [\alpha_1,\alpha_2, \cdots, \alpha_n]$ consisting of $n$ projections $\{P^k_{\alpha_k}(t_k)\}_{k=1}^n$ at times $t_1 < t_2 < \cdots < t_n$. 
Here the subscript $\alpha_k$ allows for different types of projections, for example, a position projection at $t_1$ and a momentum projection at $t_2$.
Then the decoherence functional is defined as~\cite{halliwell1994review, dowker1992quantum}
\begin{equation}
D([\alpha], [\alpha']) = \Tr[P^n_{\alpha_n}(t_n)\cdots P^1_{\alpha_1}(t_1) \rho P^1_{\alpha'_1}(t_1) \cdots P^n_{\alpha'_n}(t_n)],
\end{equation}
where 
\begin{equation}
P^k_{\alpha_k}(t_k) = e^{i(t_k-t_0)H} P^k_{\alpha_k} e^{-i(t_k-t_0)H}.
\end{equation}
It is important in consistent histories because probabilities can be assigned to histories when the decoherence functional is diagonal. 
It is easy to check that
\begin{align}
& D([\alpha], [\alpha']) = D([\alpha'], [\alpha])^*,\\
& \sum_{[\alpha]} \sum_{[\alpha']} D([\alpha], [\alpha']) = \Tr \rho = 1.
\end{align}
The diagonal elements are the probabilities for the histories $(\rho, t_0) \rightarrow (\alpha_1, t_1) \rightarrow \cdots \rightarrow (\alpha_n, t_n)$:
\begin{align}
p(\alpha_1, \alpha_2, \dots, \alpha_n)&= D(\alpha_1,\alpha_2, \dots, \alpha_n | \alpha_1,\alpha_2, \dots, \alpha_n) \nonumber \\
 &= D([\alpha], [\alpha])
\end{align}

Until now, we considered fine-grained projections $P^k_{\alpha_k}$ for fine-grained histories. The coarse-grained histories are characterised by the coarse-grained projections $\bar{P}^k_{\bar{\alpha}_k}$. To satisfy the probability sum rules, the probability for a coarse-grained history is the sum of the probabilities for its fine-grained histories. That is,
\begin{equation}
p(\bar{\alpha}_1, \bar{\alpha}_2, \dots, \bar{\alpha}_n) = \sum_{[\alpha]\in[\bar{\alpha}]} p(\alpha_1, \alpha_2, \dots, \alpha_n),
\end{equation}
where
\begin{equation}
\sum_{[\alpha]\in[\bar{\alpha}]} = \sum_{\alpha_1\in\bar{\alpha}_1}\sum_{\alpha_2\in\bar{\alpha}_2} \cdots \sum_{\alpha_n\in\bar{\alpha}_n}.
\end{equation}
On the other hand, we gain the decoherence functional for coarse-grained histories by directly summing over the fine-grained projections as~\cite{dowker1992quantum}
\begin{equation}
D([\bar{\alpha}], [\bar{\alpha}']) = \sum_{[\alpha]\in[\bar{\alpha}]} \sum_{[\alpha']\in[\bar{\alpha'}]} D([\alpha], [\alpha']).
\end{equation}
For the diagonal terms,
\begin{align}
D([\bar{\alpha}], [\bar{\alpha}]) = & \sum_{[\alpha]\in[\bar{\alpha}]} D([\alpha], [\alpha]) \nonumber \\
&+ \sum_{[\alpha]\neq [\alpha'], [\alpha]\in[\bar{\alpha}]} \sum_{[\alpha']\in[\bar{\alpha'}]} D([\alpha], [\alpha']),
\end{align}
where $[\alpha] \neq [\alpha']$ means $\alpha_k \neq \alpha'_k$ for at least one $k$.

To obey the probability sum rules that all probabilities are non-negative and summed to $1$, the sufficient and necessary condition is 
\begin{align}\label{consistencycond}
\Re[ D(\alpha_1,\alpha_2, \dots, \alpha_n | \alpha'_1,\alpha'_2, \dots, \alpha'_n) ] \nonumber \\
= p(\alpha_1, \alpha_2, \dots, \alpha_n) \delta_{\alpha_1\alpha'_1} \cdots \delta_{\alpha_n\alpha'_n}.
\end{align}
Eqn.~\eqref{consistencycond} is called the consistency condition or decoherence condition. Sets of histories obeying the condition are referred to consistent histories or decoherent histories. A stronger version of consistency condition is
\begin{align}\label{strongconsistencycond}
D(\alpha_1,\alpha_2, \dots, \alpha_n | \alpha'_1,\alpha'_2, \dots, \alpha'_n) \nonumber \\
= p(\alpha_1, \alpha_2, \dots, \alpha_n) \delta_{\alpha_1\alpha'_1} \cdots \delta_{\alpha_n\alpha'_n}.
\end{align}

The decoherence functional has a path integral representation~\cite{dowker1992quantum}. With configuration space variables $q^i(t)$ and the action $S[q^i]$, 
\begin{align}\label{pich}
D([\alpha],& [\alpha']) = \int_{[\alpha]} \mathcal{D} q^i \int_{[\alpha']} \mathcal{D} q^{i'} \nonumber \\
&\exp(iS[q^i] - iS[q^{i'}]) \delta(q_f^i - q_f^{i'}) \rho(q_0^i, q_0^{i'}),
\end{align}
where the two paths $q^i(t)$, $q^{i'}(t)$ begin at $q^i_0$, $q^{i'}_0$ respectively at $t_0$ and end at $q^i_f = q^{i'}_f$ at $t_f$, and correspond to the projections $P^k_{\alpha_k}$, $P^k_{\alpha'_k}$ made at time $t_k$ ($k = 1, 2, \dots n$).

\subsection{Temporal correlations in terms of decoherence functional}

The relation with the $n$-qubit pseudo-density matrix is arguably obvious. For example, consider an $n$-qubit pseudo-density matrix as a single qubit evolving at $n$ times. For each event, we make a single-qubit Pauli measurement $\sigma_{i_k}$ at the time $t_k$. We can separate the measurement $\sigma_{i_k}$ into two projection operators $P_{i_k}^{+1} = \frac{1}{2}(I + \sigma_{i_k} )$ and $P_{i_k}^{-1} = \frac{1}{2}(I - \sigma_{i_k} )$ with its outcomes $\pm1$. 
Corresponding to the history picture, each pseudo-density event with the measurement $\sigma_{i_k}$ corresponds to two history events with projections $P_{i_k}^{\alpha_k} (\alpha_k = \pm 1)$.
A pseudo-density matrix is built upon measurement correlations $\langle \{\sigma_{i_k}\}_{k=1}^n \rangle$. Theses correlations can be given in terms of decoherence functionals as 
\begin{align}\label{qubitch}
\langle \{\sigma_{i_k}\}_{k=1}^n \rangle & = \sum_{\alpha_1, \dots, \alpha_n} \alpha_1\cdots \alpha_n \nonumber \\
\times &\Tr[ P_{i_n}^{\alpha_n} U_{n-1} \cdots U_1 P_{i_1}^{\alpha_1} \rho P_{i_1}^{\alpha_1} U_1^{\dag} \cdots U_{n-1}^{\dag} P_{i_n}^{\alpha_n} ] \nonumber\\
& = \sum_{\alpha_1, \dots, \alpha_n} \alpha_1\cdots \alpha_n p(\alpha_1, \dots, \alpha_n) \nonumber\\
& =  \sum_{\alpha_1, \dots, \alpha_n} \alpha_1\cdots \alpha_n D([\alpha], [\alpha]),
\end{align}
where $D([\alpha], [\alpha])$ is the diagonal terms of decoherence functional with $[\alpha] = [\alpha_1, \dots, \alpha_n]$. Note that here only diagonal decoherence functionals are taken into account, which coincides with the consistency condition.

Similar relations hold for the Gaussian spacetime states. For each event, we make a single-mode quadrature measurement $\hat{q}_k$ or $\hat{p}_k$ at time $t_k$. We can separate the measurement $\hat{x}_k = \int x_k \ket{x_k}\bra{x_k} \textrm{d}x_k$ into projection operators $\ket{x_k}\bra{x_k}$ with outcomes $x_k$. Then each Gaussian event with the measurement $\hat{x}_k$ corresponds to infinite and continuous history events with projections $\ket{x_k}\bra{x_k}$.
\begin{align}\label{gaussianch}
\langle \{x_k\}_{k=1}^n \rangle & = \int_{-\infty}^{\infty}\cdots\int_{-\infty}^{\infty} \textrm{d}x_1 \cdots \textrm{d}x_n x_1\cdots x_n \nonumber \\
& \qquad \qquad \Tr[ \ket{x_n}\bra{x_n}U_{n-1} \cdots U_1 \ket{x_1}\bra{x_1} \rho \nonumber \\
& \qquad \qquad \qquad \ket{x_1}\bra{x_1} U_1^{\dag} \cdots U_{n-1}^{\dag} \ket{x_n}\bra{x_n} ]\nonumber \\
& = \int_{-\infty}^{\infty}\cdots\int_{-\infty}^{\infty} \textrm{d}x_1 \cdots \textrm{d}x_n x_1\cdots x_n p(x_1, \dots, x_n) \nonumber\\
& =  \int_{-\infty}^{\infty}\cdots\int_{-\infty}^{\infty} \textrm{d}x_1 \cdots \textrm{d}x_n x_1\cdots x_n D([x], [x]),
\end{align}
where $D([x], [x])$ is the diagonal terms of decoherence functional with $[x] = [x_1, \dots, x_n]$. 

For general spacetime states for continuous variables, we make a single-mode measurement $T(\alpha_k)$ at time $t_k$ for each event. It separates into two projection operators $P^{+1}(\alpha_k)$ and $P^{-1}(\alpha_k)$, then it follows as the $n$-qubit case.

The interesting part is to apply the lessons from consistent histories to the generalised pseudo-density matrix formulation with general measurements. 
We have argued that the spacetime density matrix can be expanded diagonally in terms of position measurements as 
\begin{align}\label{stdmfromposition}
\rho = \int_{-\infty}^{\infty}&\cdots\int_{-\infty}^{\infty} \textrm{d}x_1\cdots\textrm{d}x_n \nonumber \\
& p(x_1, \cdots, x_n) \ket{x_1}\bra{x_1} \otimes  \cdots \otimes \ket{x_n}\bra{x_n}.
\end{align}
It reminds us of the diagonal terms of the decoherence functional. It is possible to build a spacetime density matrix from all possible decoherence functionals as 
\begin{align}
\rho = \int_{-\infty}^{\infty}&\cdots\int_{-\infty}^{\infty} \textrm{d}x_1\textrm{d}x'_1\cdots\textrm{d}x_n\textrm{d}x'_n \nonumber \\
&D(x_1, \dots, x_n | x'_1, \dots x'_n) \ket{x_1}\bra{x'_1} \otimes  \cdots \otimes \ket{x_n}\bra{x'_n}.
\end{align}
Applying the strong consistency condition to the above equation, we gain Eqn.~\eqref{stdmfromposition} again. 
This argues why it is effective to only consider diagonal terms in position measurements. which is originally taken for convenience. 

Similarly, the spacetime Wigner function from weak measurements is easily taken as a generalisation for the diagonal terms of the decoherence functional allowing for general measurements. 
Recall that a generalised effect-valued measure is represented by 
\begin{equation}
\hat{f}(q, p) = C\exp\left[ -\alpha[(\hat{q}-q)^2 + \lambda (\hat{p}-p)^2] \right].
\end{equation}
The generalised decoherence functional for weak measurements is then given by
\begin{equation}
D(q, p, q', p', \tau | \hat{\rho}) = \Tr \left[ \mathcal{F}(q ,p, q', p'; \tau)  \hat{\rho} \right],
\end{equation}
where
\begin{align}
&\mathcal{F}(q ,p, q',p'; \tau)  \hat{\rho} \nonumber \\
= & \int \textrm{d}\mu_G [q(t), p(t)] \int \textrm{d}\mu_G [q'(t), p'(t)] \delta \left( q - \frac{1}{\tau}\int_0^{\tau} \textrm{d}t q(t) \right) \nonumber \\
& \times \delta \left( p - \frac{1}{\tau}\int_0^{\tau} \textrm{d}t p(t) \right) \delta \left( q' - \frac{1}{\tau}\int_0^{\tau} \textrm{d}t q'(t) \right) \nonumber \\
& \times \delta \left( p' - \frac{1}{\tau}\int_0^{\tau} \textrm{d}t p'(t) \right) \exp[-\frac{i}{\hbar}\hat{H}\tau] \times \nonumber \\
&\mathcal{T} \exp \left[ -\frac{\gamma}{2} \int_{0}^{\tau} \textrm{d}t [(\hat{q}_H(t) - q(t))^2 + \lambda (\hat{p}_H(t) - p(t))^2 ] \right] \hat{\rho} \nonumber \\
& \mathcal{T}^* \exp \left[ -\frac{\gamma}{2} \int_{0}^{\tau} \textrm{d}t [(\hat{q'}_H(t) - q'(t))^2 + \lambda (\hat{p'}_H(t) - p'(t))^2 ] \right] \nonumber \\
& \times \exp[\frac{i}{\hbar}\hat{H}\tau],
\end{align}
here
\begin{align}
&\textrm{d}\mu_G [q(t), p(t)] = \lim_{N \rightarrow \infty} \left( \frac{\gamma\tau\sqrt{\lambda}}{\pi N} \prod_{s=1}^{N} \textrm{d}q(t_s) \textrm{d}p(t_s) \right),\\
&\textrm{d}\mu_G [q'(t), p'(t)] = \lim_{N \rightarrow \infty} \left( \frac{\gamma\tau\sqrt{\lambda}}{\pi N} \prod_{s=1}^{N} \textrm{d}q'(t_s) \textrm{d}p'(t_s) \right),
\end{align}
and
\begin{align}
&\hat{q}_H(t) = \exp\left[ \frac{i}{\hbar} \hat{H} t \right] \hat{q} \exp\left[ -\frac{i}{\hbar} \hat{H} t \right], \nonumber \\
&\hat{q'}_H(t) = \exp\left[ \frac{i}{\hbar} \hat{H} t \right] \hat{q'} \exp\left[ -\frac{i}{\hbar} \hat{H} t \right], \nonumber \\
&\hat{p}_H(t) = \exp\left[ \frac{i}{\hbar} \hat{H} t \right] \hat{p} \exp\left[ -\frac{i}{\hbar} \hat{H} t \right], \nonumber \\
&\hat{p'}_H(t) = \exp\left[ \frac{i}{\hbar} \hat{H} t \right] \hat{p'} \exp\left[ -\frac{i}{\hbar} \hat{H} t \right].
\end{align}
The diagonal terms under the strong consistency condition reduce to the form in Ref
~\cite{zhang2020different}: 
\begin{equation}\label{stwffromweak}
p(q, p, \tau | \hat{\rho}) = \Tr \mathcal{F}(q ,p; \tau)  \hat{\rho},
\end{equation}
where
\begin{align}\label{stweak}
&\mathcal{F}(q ,p; \tau)  \hat{\rho} \nonumber \\
= & \int \textrm{d}\mu_G [q(t), p(t)] \delta \left( q - \frac{1}{\tau}\int_0^{\tau} \textrm{d}t q(t) \right) \nonumber \\
&\times \delta \left( p - \frac{1}{\tau}\int_0^{\tau} \textrm{d}t p(t) \right) \exp[-\frac{i}{\hbar}\hat{H}\tau] \times \nonumber \\
& \mathcal{T} \exp \left[ -\frac{\gamma}{2} \int_{0}^{\tau} \textrm{d}t [(\hat{q}_H(t) - q(t))^2 + \lambda (\hat{p}_H(t) - p(t))^2 ] \right] \hat{\rho} \nonumber \\
& \mathcal{T}^* \exp \left[ -\frac{\gamma}{2} \int_{0}^{\tau} \textrm{d}t [(\hat{q}_H(t) - q(t))^2 + \lambda (\hat{p}_H(t) - p(t))^2 ] \right] \nonumber \\
&\times \exp[\frac{i}{\hbar}\hat{H}\tau].
\end{align}

Now we conclude the relation between decoherence functionals in consistent histories and temporal correlations in pseudo-density matrices. 	
\begin{claim}
The decoherence functional in consistent histories is the probabilities in temporal correlations of pseudo-density matrices, e.g., as in Eqn.~\eqref{qubitch} and Eqn.~\eqref{gaussianch}.
\end{claim}
Thus, we establish the relationship between consistent histories and all possible forms of pseudo-density matrix. 
From the consistency condition, we also have a better argument for why spacetime states for general measurements are defined in the diagonal form. It is not a coincide.



\section{Generalised non-local games}
Game theory studies mathematical models of competition and cooperation under strategies among rational decision-makers~\cite{myerson2013game}. Here we give an introduction to nonlocal games, quantum-classical nonlocal games, and quantum-classical signalling games. Then we show the relation between quantum-classical signalling games and pseudo-density matrices, and comment on the relation between general quantum games and indefinite causal order. 

\subsection{Introduction to non-local games}
The interests for investigating non-local games start from interactive proof systems with two parties, the provers and the verifiers. They exchange information to verify a mathematical statement. A nonlocal game is a special kind of interactive proof system with only one round and at least two provers who play in cooperation against the verifier. In nonlocal games, we refer to the provers as Alice, Bob, $\dots$, and the verifier as the referee. In Ref.~\cite{cleve2004consequences}, nonlocal games were formally introduced with shared entanglement and used to formulate the CHSH inequality~\cite{clauser1969proposed}. Here we introduce the CHSH game as an example and then give the general form of a non-local game. 

The CHSH game has two cooperating players, Alice and Bob, and a referee who asks questions and collects answers from the players. The basic rules of the CHSH game are as the following: 

1) There are two possible questions $x \in \{0, 1\}$ for Alice and two possible questions $y \in \{0, 1\}$ for Bob. Each question has an equal probability as $p(x, y) = \frac{1}{4}, \forall x, \forall y$. 

2) Alice answers $a \in \{0, 1\}$ and Bob $b \in \{0, 1\}$. 

3) Alice and Bob cannot communicate with each other after the game begins. 

4) If $a \oplus b = x \cdot y$, then they win the game, otherwise they lose. 

For a classical strategy, that is, Alice and Bob use classical resources, they win with the probability at most $\frac{3}{4}$. 
Alice and Bob can also adopt a quantum strategy. If they prepare and share a joint quantum state $\ket{\Phi^+} = \frac{1}{\sqrt{2}}(\ket{00}+\ket{11})$ and make local measurements based on the questions they receive separately, then they can achieve a higher winning probability $\cos^2(\pi/8) \approx 0.854$. 

In general, a non-local game $G$ is formulated by $(\pi, l)$ on 
\begin{equation}
\overrightarrow{nl} = \langle \mathcal{X}, \mathcal{Y}; \mathcal{A}, \mathcal{B}; l \rangle,
\end{equation}
where $\mathcal{X}$, $\mathcal{Y}$ are question spaces of Alice and Bob and $\mathcal{A}$, $\mathcal{B}$ are answer spaces of Alice and Bob.
Here $\pi(x, y)$ is a probability distribution of the question spaces for Alice and Bob in the form
$\pi: \mathcal{X} \times \mathcal{Y} \rightarrow [0, 1]$, and 
$l(a, b | x, y)$ is a function of question and answer spaces for Alice and Bob to decide whether they win or lose in the form
$l:  \mathcal{X} \times \mathcal{Y} \times \mathcal{A} \times \mathcal{B} \rightarrow [0, 1];$
for example, if they win, $l = 1$; otherwise lose with $l = 0$. 
For any strategy, the probability distribution for answers $a, b$ of Alice and Bob given questions $x, y$, respectively, is referred to as the correlation function $p(a, b | x, y)$ of the form 
\begin{equation}
p:  \mathcal{X} \times \mathcal{Y} \times \mathcal{A} \times \mathcal{B} \rightarrow [0, 1].
\end{equation}
with the condition
$\sum_{a,b} p(a, b | x, y) = 1.$
With a classical source,
\begin{equation}
p_c(a, b | x, y) = \sum_{\lambda}  \pi(\lambda) d_A(a | x, \lambda) d_B(b | y, \lambda),
\end{equation}
where $d_A(a | x, \lambda)$ is the probability of answering $a$ given the parameter $\lambda$ and the question $b$ and similar for $d_B(b | y, \lambda)$;
with a quantum source,
\begin{equation}
p_q(a, b | x, y) = \Tr [\rho_{AB} (P_A^{a|x} \otimes Q_B^{b|y})],
\end{equation}
where $\rho_{AB}$ is the quantum state shared by Alice and Bob, $P_A^{a|x}$ is the measurement made by Alice with the outcome $a$ given $x$, $Q_B^{b|y}$ is the measurement made by Bob with the outcome $b$ given $y$. 
Then the optimal winning probability is given by 
\begin{equation}
\mathbb{E}_{\overrightarrow{nl}} [*] \equiv \max \sum_{x,y} \pi(x, y) \sum_{a,b} l(a, b | x, y) p_{c/q}(a, b | x, y).
\end{equation}

\subsection{Quantum-classical non-local \& signalling games}

First we introduce a generalised version of non-local games where the referee asks quantum questions instead of classical questions (therefore this type of non-local games are refereed to quantum-classical)~\cite{buscemi2012all}. Then we give the temporal version of these quantum-classical non-local games as quantum-classical signalling games~\cite{rosset2018resource}. 

\subsubsection{Quantum-classical non-local games}
We now recap the model of quantum-classical non-local games~\cite{buscemi2012all}, in which the questions are quantum rather than classical. More specifically, the referee sends quantum registers to Alice and Bob instead of classical information. 

For a non-local game, with the question spaces $\mathcal{X} =  \{ x \} $ and $\mathcal{Y} =  \{ y \}$, the referee associates two quantum ancillary systems $X$ and $Y$ such that $\dim \mathcal{H}_{X} \geq |\mathcal{X}|$, $\dim \mathcal{H}_{Y} \geq |\mathcal{Y}|$, the systems are in the states $\tau^{x}_{X} = \ket{x}\bra{x}$ and $\tau^{y}_{Y} = \ket{y}\bra{y}$ with the questions $x \in \mathcal{X}$ and $y \in \mathcal{Y}$. 
Assume that Alice and Bob share a quantum state $\rho_{AB}$. 
Given the answer sets $\mathcal{A} = \{ a \}$ and $\mathcal{B} = \{ b \}$ and quantum systems $XA$ and $YB$, Alice and Bob can make the corresponding POVMs $P^a_{XA}$ and $Q^b_{YB}$ in the linear operators on the Hilbert space $\mathcal{H}_{XA}$ and $\mathcal{H}_{YB}$, such that $\sum_a P^a_{XA} = \mathbbm{1}_{XA}$ and $\sum_b Q^b_{YB} = \mathbbm{1}_{YB}$. 
Then the probability distribution for the questions and answers of Alice and Bob, that is, the correlation function $P(a, b | x, y)$, is given by
\begin{equation}
P(a, b | x, y) = \Tr [(P^a_{XA} \otimes Q^b_{YB}) (\tau^{x}_{X} \otimes \rho_{AB} \otimes \tau^{y}_{Y}) ].
\end{equation}

Quantum-classical non-local games replace classical inputs with quantum ones, formulated by $(\pi(x, y), l(a, b | x, y))$ on 
\begin{equation}
\overrightarrow{qcnl} = \langle \{\tau^x\}, \{\omega^y\}; \mathcal{A}, \mathcal{B}; l \rangle.
\end{equation}
The referee picks $x \in \mathcal{X}$ and $y \in \mathcal{Y}$ with the probability distribution $\pi(x, y)$ as the classical-classical non-local game. 
With a classical source,
\begin{equation}
p_c(a, b | x, y) = \sum_{\lambda} \pi(\lambda) \Tr [ (\tau^x_X \otimes \omega^y_Y) (P_X^{a|\lambda} \otimes Q_Y^{b|\lambda})];
\end{equation}
with a quantum source,
\begin{equation}
p_q(a, b | x, y) = \Tr [ (\tau^x_X \otimes \rho_{AB} \otimes \omega^y_Y)  (P_{XA}^{a} \otimes Q_{BY}^{b})].
\end{equation}
The optimal winning probability is, again, given by
\begin{equation}
\mathbb{E}_{\overrightarrow{qcnl}} [*] \equiv \max \sum_{x,y} \pi(x, y) \sum_{a,b} l(a, b | x, y) p_{c/q}(a, b | x, y).
\end{equation}

\subsubsection{Quantum-classical signalling games}

In quantum-classical signalling games~\cite{rosset2018resource}, instead of two players Alice and Bob, we consider only one player Abby at two successive instants in time.
Then quantum-classical signalling games change the Alice-Bob duo to a timelike structures of single player Abby with
\begin{equation}
\overrightarrow{qcsg} = \langle \{\tau^x\}, \{\omega^y\}; \mathcal{A}, \mathcal{B}; l \rangle.
\end{equation}
With unlimited classical memory,
\begin{equation}
p_c(a, b | x, y) = \sum_{\lambda} \pi(\lambda) \Tr [ \tau^x_X P_X^{a|\lambda} ] \Tr [ \omega^y_Y Q_Y^{b|a, \lambda}].
\end{equation}
For admissible quantum strategies, suppose Abby at $t_1$ receives $\tau^x_X$ and makes a measurement of instruments $\{\Phi_{X\rightarrow A}^{a|\lambda}\}$, and gains the outcome $a$. Then the quantum output goes through the quantum memory $\mathcal{N}: A \rightarrow B$. The output of the memory and $\omega^y_Y$ received by Abby at $t_2$ are fed into a measurement $\{\Psi_{BY}^{b|a, \lambda}\}$, with outcome b. 
Then
\begin{align}
&p_q(a, b | x, y)\nonumber \\
 =  &\sum_{\lambda} \pi(\lambda) \Tr [ (\{ (\mathcal{N}_{A \rightarrow B} \circ \Phi_{X \rightarrow A}^{a | \lambda})(\tau^x_X) \} \otimes \omega^y_Y)\Psi_{BY}^{b | a, \lambda}].
\end{align}
The optimal payoff function is, again, given by
\begin{equation}
\mathbb{E}_{\overrightarrow{qcsg}} [*] \equiv  \max \sum_{x,y} \pi(x, y) \sum_{a,b} l(a, b | x, y) p_{c/q}(a, b | x, y).
\end{equation}

\subsection{Temporal correlations from signalling games}

To compare quantum-classical signalling games with pseudo-density matrices, first we generalise the finite-dimensional pseudo-density matrices from Pauli measurements to general positive-operator valued measures(POVMs). Recall that a POVM is a set of Hermitian positive semi-definite operator $\{E_i\}$ on a Hilbert space $\mathcal{H}$ which sum up to the identity $\sum_i E_i = \mathbbm{1}_{\mathcal{H}}$. Instead of making a single-qubit Pauli measurement at each event, we make a measurement $E_i = M_i^{a \dag}M_i^a$ with the outcome $a$. For each event, there is a measurement $\mathcal{M}_i: \mathcal{L}(\mathcal{H}^X) \rightarrow \mathcal{L}(\mathcal{H}^A), \tau^x_X \mapsto \sum_i M_i^a \tau^x_X M_i^{a \dag}$ with $\sum M_i^{a \dag} M_i^a = \mathbbm{1}_{\mathcal{H}^X}$.

Now we map the generalised pseudo-density matrices to quantum-classical signalling games. Assume $\omega_Y^y$ to be trivial. For Abby at the initial time and the later time, we consider $\Phi_{X \rightarrow A}^{a}: \tau^x_X \rightarrow \sum_i M_i^a \tau^x_X M_i^{a \dag}$, $\sum M_i^{a \dag} M_i^a = \mathbbm{1}_{\mathcal{H}^A}$. 
Between two times, the transformation from $A$ to $B$ is given by $\mathcal{N}: \rho_A \rightarrow \sum_j N_j \rho_A N_j^{\dag}$ with $\sum_j N_j^{\dag}N_j = \mathbbm{1}_{\mathcal{H}^A}$. Then 
\begin{align}
p_q(a, b | x, y) & =  \Tr [ \{ (\mathcal{N}_{A \rightarrow B} \circ \Phi_{X \rightarrow A}^{a})(\tau^x_X) \} \Psi_{B}^{b | a}] \nonumber \\
& =  \sum_{ik} \Tr [ \mathcal{N} \{ M_i^{a} \tau^x_X M_i^{a \dag} \} \Psi_{B}^{b | a}] \nonumber \\
& = \sum_{ijk} \Tr [ N_j M_i^{a} \tau^x_X M_i^{a \dag} N_j^{\dag} \Psi_{B}^{b | a}] \\
 \langle \{ \Phi, \Psi \} \rangle & = \sum_{a, b}  ab p_q(a, b | x, y)
\end{align}
It is the temporal correlation given by pseudo-density matrices. 
That is, a quantum-classical signalling game with a trivial input at later time corresponds to a pseudo-density matrix with quantum channels as measurements. 
\begin{claim}
The probability in a quantum-classical signalling game with a trivial input at the later time corresponds to the probability in a pseudo-density matrix with quantum channels as measurements. 
\end{claim}

It is also convenient to establish the relation between generalised games in time and indefinite causal structures with double Hilbert spaces for each event. For completeness, we also mention that Gutoski and Watrous~\cite{gutoski2007toward} proposed a general theory of quantum games in terms of the Choi-Jamio\l{}kowski representation, which is an equivalent formulation of indefinite causal order. 

\section{Out-of-time-order correlations (OTOCs)}
In this section we introduce out-of-time-order correlation functions, find a simple method to calculation these temporal correlations via the pseudo-density matrix formalism.

\subsection{Brief introduction to OTOCs}

Consider local operators $W$ and $V$. With a Hamiltonian $H$ of the system, the Heisenberg representation of the operator $W$ is given as $W(t) = e^{iHt}We^{-iHt}$. 
Out-of-time-order correlation functions (OTOCs)~\cite{maldacena2015bound, roberts2016chaos} are usually defined as 
\begin{equation}
\langle VW(t)V^{\dag}W^{\dag}(t) \rangle = \langle VU(t)^{\dag}WU(t)V^{\dag}U^{\dag}(t)W^{\dag}U(t)], 
\end{equation}
where $U(t) = e^{-iHt}$ is the unitary evolution operator and the correlation is evaluated on the thermal state $\langle \cdot \rangle = \Tr[e^{-\beta H} \cdot]/ \Tr[e^{-\beta H}]$. 
Note that OTOC is usually defined for the maximally mixed state $\rho = \frac{\mathbbm{1}}{d}$. 
Consider a correlated qubit chain. Measure $V$ at the first qubit and $W$ at the last qubit. Since the chain is correlated in the beginning, we have OTOC as $1$ at the early time. As time evolves and the operator growth happens, OTOC will approximate to $0$ at the late time.


\subsection{Calculating OTOCs via pseudo-density matrices}
In this subsection we make a connection between OTOCs and pseudo-density matrix formalism. If we consider a qubit evolving in time and backward, we can get a tripartite pseudo-density matrix. 
In particular, we consider measuring $A$ at $t_1$, $B$ at $t_2$ and $A$ again at $t_3$ and assume the evolution forwards is described by $U$ and backward $U^{\dag}$. 
Then the probability is given by 
\begin{equation}\label{otocrel}
\Tr[A U^{\dag} B U A \rho A^{\dag} U^{\dag} B^{\dag} U A^{\dag}] = \Tr[A B(t) A \rho A^{\dag} B^{\dag}(t) A^{\dag}] 
\end{equation}
If we assume that $AA^{\dag} = A$, $\rho = \frac{\mathbbm{1}}{d}$, Eqn.~\eqref{otocrel} will reduce to the OTOC. 
\begin{claim}
OTOCs can be represented as temporal correlations in pseudo-density matrices with half the numbers of steps for calculation; for example, a four-point OTOC, usually calculated by evolving forwards and backwards twice, is represented by a tripartite pseudo-density matrix with only once evolving forwards and backwards. 
\end{claim}

\section{A unified picture}

Now we consider a unified picture in which temporal correlations serve as a connection for indefinite causal order, consistent histories, generalised quantum games and OTOCs. Given a tripartite pseudo-density matrix, a qubit in the state $\rho$ evolves in time under the unitary evolution $U$ and then back in time under $U^{\dag}$. The correlations in the pseudo-density matrix are given as 
\begin{equation}\label{tripdm}
\langle \sigma_i, \sigma_j, \sigma_k \rangle = \sum_{\alpha, \beta, \gamma = \pm 1} \alpha \beta \gamma \Tr[P^{\gamma}_k U^{\dag} P^{\beta}_j U P^{\alpha}_i \rho P^{\alpha}_i U^{\dag} P^{\beta}_j U]	
\end{equation}
where $P^{\alpha}_i = \frac{1}{2}(\mathbb{I}+\alpha\sigma_i)$, $P^{\beta}_j = \frac{1}{2}(\mathbb{I}+\beta\sigma_j)$ and $P^{\gamma}_k = \frac{1}{2}(\mathbb{I}+\gamma\sigma_k)$. 
As the pseudo-density matrix belongs to indefinite causal order, we won't discuss the transform for indefinite causal order. 

For consistent histories, we assume the state in $\rho$ at the initial time and construct a set of histories $[\chi] = [\alpha \rightarrow \beta \rightarrow \gamma]$ with projections $\{P^{\alpha}_i, P^{\beta}_j, P^{\gamma}_k\}$. Then the decoherence functional is given as 
\begin{equation}
D([\xi], [\xi']) =\Tr[P^{\gamma}_k U^{\dag} P^{\beta}_j U P^{\alpha}_i \rho P^{\alpha'}_i U^{\dag} P^{\beta'}_j U P^{\gamma'}_k ] 	
\end{equation}
When we apply the consistency conditions, it is part of Eqn.~\eqref{tripdm} as 
\begin{equation}
D([\xi], [\xi]) =\Tr[P^{\gamma}_k U^{\dag} P^{\beta}_j U P^{\alpha}_i \rho P^{\alpha}_i U^{\dag} P^{\beta}_j U P^{\gamma}_k ],
\end{equation}
\begin{equation}
\langle \sigma_i, \sigma_j, \sigma_k \rangle = \sum_{\alpha, \beta, \gamma = \pm 1} \alpha \beta \gamma D([\xi], [\xi]). 
\end{equation}

A quantum-classical signalling game is described in terms of one player Abby at two times in a loop, or one player Abby at three times with evolution $U$ and $U^{\dag}$. 
The quantum-classical signalling game is formulated by $(\pi(x,y), l(a, b|x, y))$ on 
\begin{equation}
\overrightarrow{qcsg} = \langle \{\tau^x\}, \{\omega^y\}, \{\eta^z\}; \mathcal{A}, \mathcal{B}, \mathcal{C}; l \rangle.
\end{equation}
The referee associates three quantum systems in the states $\tau^x$, $\omega^y$ and $\eta^z$ with the questions chosen from the question spaces $x \in \mathcal{X}$, $y \in \mathcal{Y}$, and $z \in \mathcal{Z}$. Suppose Abby at $t_1$ receives $\tau^x_X$ and makes a measurement of instruments $\{M_i^a\}_i$ with the outcome $a$. From $t_1$ to $t_2$, the quantum output evolves under the unitary quantum memory $U: A \rightarrow B$. After that, Abby receives the output of the channel and $\omega^y$, and makes a measurement of instruments $\{N_j^b\}_j$ with the outcome $b$. Then, we can consider that either the quantum memory goes backwards to $t_1$ or evolves under $U^{\dag}: B \rightarrow C$ to $t_3$. Abby receives the output of the channel again and $\eta^z$, and makes a measurement of instruments $\{O_k^c\}_k$ with the outcome $c$.
Then we have 
\begin{align}
&p_q(a, b, c | x, y, z)\nonumber \\
= & \sum_{\lambda, i, j, k} \pi(\lambda) \Tr[O_k^c U^{\dag} N_j^b U M_i^a \rho M_i^a U^{\dag} N_j^b U O_k^c].
\end{align}
If we properly choose the measurements, we will have the decoherence functionals and the probabilities in the correlations of pseudo-density matrix.

What is more, the tripartite pseudo-density matrix we describe is just the one we used to construct OTOC. 
Thus, through this tripartite pseudo-density matrix, we gain a unified picture for indefinite causal order, consistent histories, generalised quantum games and OTOCs in which temporal correlations are the same or operationally equivalent. Thus all these approaches are mapping into each other directly in this particular case via temporal correlations. Generalisation to more complicated scenarios are straightforward.  

\section{Path integrals}

The path integral approach~\cite{feynman2010quantum} is a representation of quantum theory, not only useful in quantum mechanics but also quantum statistical mechanics and quantum field theory. It generalises the action principle of classical mechanics and one computes a quantum amplitude by replacing a single classical trajectory with a functional integral of infinite numbers of possible quantum trajectories. 
Here we argue that the path integral approach of quantum mechanics use amplitude as the measure in correlation functions rather than probability measure in the above formalisms, and thus treats temporal correlations in a qualitatively different way.

\subsection{Introduction to path integrals}
Now we briefly introduce path integrals and correlation functions~\cite{zinn2010path}. Here we follow the Euclidean path integrals in statistical mechanics for convenience to illustrate the example of harmonic oscillators in the next subsection. 
Consider a bound operator in a Hilbert space $U(t_2, t_1) (t_2 \geq t_1)$ as the evolution from time $t_1$ to $t_2$, which satisfies the Markov property in time as
\begin{equation}
U(t_3, t_2)U(t_2, t_1) = U(t_3, t_1), \forall\   t_3 \geq t_2 \geq t_1 \qquad U(t, t) = \mathbbm{1}.
\end{equation}
We further assume that $U(t, t')$ is differentiable and the derivative is continuous:
\begin{equation}
\left.\frac{\partial U(t, t')}{\partial t}\right	|_{t=t'} = -H(t)/\hbar
\end{equation}
where $\hbar$ is a real parameter, and later identified with Planck's constant; $H = i \tilde{H}$ where $\tilde{H}$ is the quantum Hamiltonian. 
Then
\begin{equation}
U(t'', t') = \prod_{m=1}^n U[t'+m\epsilon, t'+(m-1)\epsilon], \qquad n\epsilon = t"-t'.
\end{equation}
The position basis for $\hat{q}\ket{q} = q\ket{q}$ is orthogonal and complete: $\langle q' \ket{q} = \delta(q-q')$, $\int \textrm{d}q \ket{q}\bra{q} = \mathbbm{1}$. 
We have 
\begin{equation}
\bra{q''}U(t'', t')\ket{q'} = \int \prod_{k=1}^{n-1} \textrm{d}q_k \prod_{k=1}^n \bra{q_k} U(t_k, t_{k-1})\ket{q_{k-1}}
\end{equation}
with $t_k = t' + k\epsilon, q_0 = q', q_n = q''$. 
Suppose that the operator $H$ is identified with a quantum Hamiltonian of the form
\begin{equation}
H = \hat{\bm{p}}^2/2m + V(\hat{\bm{q}}, t)	
\end{equation}
where $\bm{p}, \bm{q} \in \mathbb{R}^d$. We have 
\begin{equation}
\bra{\bm{q}}U(t, t')\ket{\bm{q'}} = \left( \frac{m}{2\pi\hbar(t-t')} \right)^{d/2} \exp[-\mathcal{S}(\bm{q})/\hbar]
\end{equation}
where 
\begin{equation}
\mathcal{S}(\bm{q}) = \int_{t'}^t \textrm{d}\tau \left[\frac{1}{2}m\dot{\bm{q}}^2(\tau) + V(\bm{q}(\tau), \tau)\right] + O((t-t')^2)	,
\end{equation}
and 
\begin{equation}
\bm{q}(\tau) = \bm{q}'+\frac{\tau - t'}{t- t'}(\bm{q}-\bm{q}').
\end{equation}
We consider short time slices, then
\begin{align}
&\bra{\bm{q''}}U(t'', t')\ket{\bm{q'}} \nonumber \\
= &\lim_{n\rightarrow\infty} \left( \frac{m}{2\pi\hbar\epsilon} \right)^{dn/2} \int \prod_{k=1}^{n-1} \textrm{d}^d q_k \exp[-\mathcal{S}(\bm{q}, \epsilon)/\hbar],
\end{align}
with 
\begin{equation}\label{sss}
\mathcal{S}(\bm{q}, \epsilon) =\sum_{k=0}^{n-1} \int_{t_k}^{t_{k+1}} \textrm{d}t \left[\frac{1}{2}m\dot{\bm{q}}^2(t) + V(\bm{q}(t), t)\right] + O(\epsilon^2).
\end{equation}
Introducing a linear and continuous trajectory 
\begin{equation}
\bm{q}(t) = \bm{q}_k + \frac{t- t_k}{t_{k+1} - t_k}(\bm{q}_{k+1} - \bm{q}_k) \ \ \text{for} \ \ t_k\leq t \leq t_{k+1},
\end{equation}
we can rewrite Eqn.~\eqref{sss} as
\begin{equation}
\mathcal{S}(\bm{q}, \epsilon) = \int_{t'}^{t''} \textrm{d}t \left[\frac{1}{2}m\dot{\bm{q}}^2(t) + V(\bm{q}(t), t)\right] + O(n\epsilon^2).
\end{equation}
Taking $n \rightarrow \infty$ and $\epsilon \rightarrow 0$ with $n\epsilon = t''-t'$ fixed, we have
\begin{equation}
\mathcal{S}(\bm{q}) = \int_{t'}^{t''} \textrm{d}t \left[\frac{1}{2}m\dot{\bm{q}}^2(t) + V(\bm{q}(t), t)\right]
\end{equation}
as the Euclidean action. 
The path integral is thus defined as 
\begin{equation}
\bra{\bm{q}''}U(t'',t')\ket{\bm{q}'} = \int_{\bm{q}(t') = \bm{q}'}^{\bm{q}(t'') = \bm{q}''}[\textrm{d}\bm{q}(t)] \exp(-\mathcal{S}(\bm{q})/ \hbar),
\end{equation}
where a normalisation of $\mathcal{N} = ( \frac{m}{2\pi\hbar\epsilon} )^{dn/2}$ is hidden in $[\textrm{d}\bm{q}(t)]$. 

The quantum partition function $\mathcal{Z}(\beta) = \Tr e^{-\beta H}$ ($\beta$ is the inverse temperature) can be written in terms of path integrals as 
\begin{align}
\mathcal{Z}(\beta) & = \Tr e^{-\beta H} = \Tr U(\hbar\beta, 0) \nonumber \\
& = \int \textrm{d}q''\textrm{d}q'\delta(\bm{q}''-\bm{q}')\bra{\bm{q}''}U(\hbar\beta, 0)\ket{\bm{q}'} \nonumber\\
& = \int_{\bm{q}(0) = \bm{q}(\hbar\beta)} [\textrm{d}q(t)]\exp[-\mathcal{S}(\bm{q})/\hbar].
\end{align}
The integrand $e^{-\mathcal{S}(\bm{q})/\hbar}$ is a positive measure and defines the corresponding expectation value as 
\begin{equation}
\langle \mathcal{F}(q) \rangle = \mathcal{N}\int [\textrm{d}q(t)] \mathcal{F}(q) \exp[-\mathcal{S}(\bm{q})/\hbar],
\end{equation}
where $\mathcal{N}$ is chosen for $\langle 1 \rangle = 1$.
Moments of the measure in the form as 
\begin{align}
&\langle q(t_1)q(t_2) \cdots q(t_n) \rangle \nonumber \\
= &\mathcal{N}\int [\textrm{d}q(t)]q(t_1)q(t_2) \cdots q(t_n)\exp[-\mathcal{S}(\bm{q})/\hbar]
\end{align}
are the $n$-point correlation function. 
Suppose for the finite time interval $\beta$ periodic boundary conditions hold as $q(\beta/2) = q(-\beta/2)$. The normalisation is given as $\mathcal{N} = \mathcal{Z}^{-1}(\beta)$. 
Then we define 
\begin{equation}
Z^{(n)}(t_1, \cdots, t_n) = 	\langle q(t_1) \cdots q(t_n) \rangle.
\end{equation}
The generating functional of correlation functions is 
\begin{align}
\mathcal{Z}(f) & = \sum_{n=0} \frac{1}{n!}\int \textrm{d}t_1 \cdots \textrm{d}t_n 	Z^{(n)}(t_1, \cdots, t_n) f(t_1) \cdots f(t_n) \nonumber \\
& = \sum_{n=0} \frac{1}{n!}\int \textrm{d}t_1 \cdots \textrm{d}t_n \langle q(t_1) \cdots q(t_n) \rangle f(t_1) \cdots f(t_n) \nonumber \\
& = \left \langle \exp\left[ \int\textrm{d}t q(t)f(t)\right] \right \rangle
\end{align}
What is more, the $n$-point quantum correlation functions in time appear as continuum limits of the correlation functions of $1D$ lattice in classical statistical models.
The path integral, thus, represent a mathematical relation between classical statistical physics on a line and quantum statistical physics of a point-like particle at thermal equilibrium. 
This is the first example of the quantum-classical correspondence which maps between quantum statistical physics in $D$ dimensions and classical statistical physics in $D + 1$ dimensions~\cite{zinn2010path}.

\subsection{Temporal correlations in path integrals are different}
Here we take two-point correlations functions:
\begin{equation}\label{tpcf}
\langle q(t_1)q(t_2) \rangle  = \frac{\int [\textrm{d}q(t)]q(t_1)q(t_2)\exp[-\mathcal{S}(\bm{q})/\hbar]}{\int [\textrm{d}q(t)]\exp[-\mathcal{S}(\bm{q})/\hbar]}
\end{equation}
In the Gaussian representation of pseudo-density matrices, temporal correlation for $q_1$ at $t_1$ and $q_2$ at $t_2$ with the evolution $U$ and the initial state $\ket{q_1}$ is given as
\begin{align}
\langle \{q_1, q_2\} \rangle& = \int\textrm{d}q_1 \textrm{d}q_2 q_1 q_2 |\bra{q_2}U\ket{q_1}|^2 \nonumber \\
& =  \int\textrm{d}q_1 \textrm{d}q_2 q_1 q_2 \left| \int_{q(t_1) = q_1}^{q(t_2) = q_2} [\textrm{d}q(t)] \exp[-\mathcal{S}(\bm{q})/\hbar]\right|^2 
\end{align}
Correlations are defined as the expectation values of measurement outcomes. However, path integrals and pseudo-density matrices use different positive measure to calculate the expectation values. 
The correlations in path integrals use the amplitude as the measure, while in pseudo-density matrices the measure is the absolute square of the amplitude, or we say the probability. 

To see the difference, we consider a quantum harmonic oscillator. The Hamiltonian is given as $H = \hat{p}^2/2m + m\omega^2\hat{q}^2/2$. 
Note that the quantum amplitude of a quantum harmonic oscillator is given as
\begin{align}
\bra{q_2}&U(t_2, t_1)\ket{q_1} = \left(\frac{m\omega}{2\pi\hbar\sinh \omega \tau}	\right)^{1/2}\nonumber \\
&\times \exp\left\{ -\frac{m\omega}{2\hbar\sinh \omega\tau} [(q_1^2+q_2^2)\cosh \omega \tau - 2q_1q_2] \right\},
\end{align}
where $\tau = t_2 - t_1$.
In the Gaussian representation of pseudo-density matrices, temporal correlations are represented as
\begin{equation}
\langle \{ q_1, q_2 \} \rangle = \int\textrm{d}q_1 \textrm{d}q_2 q_1 q_2 |\bra{q_2}U\ket{q_1}|^2 = \frac{\hbar}{8m\omega\sinh^2\omega\tau}.
\end{equation}
However, in the path integral formalism, we consider 
\begin{equation}
\Tr U_G(\tau/2, -\tau/2; b) = \int [\textrm{d}q(t)] \exp[-\mathcal{S}_G(q, b)/\hbar]	
\end{equation}
with  
\begin{equation}
\mathcal{S}_G(q, b) = \int_{-\tau/2}^{\tau/2} \textrm{d}t \left[\frac{1}{2}m\dot{q}^2(t) + \frac{1}{2}m\omega^2q^2(t) - b(t)q(t)\right]
\end{equation}
and periodic boundary conditions $q(\tau/2) = q(-\tau/2)$. 
We have 
\begin{align}
\mathcal{Z}_G(\beta, b) &= \Tr U_G(\hbar\beta/2, -\hbar\beta/2; b) \nonumber \\
& = \mathcal{Z}_0(\beta) \left \langle \exp\left[ \frac{1}{\hbar}\int_{-\hbar\beta/2}^{\hbar\beta/2}\textrm{d}t b(t)q(t)\right] \right \rangle_0
\end{align}
where $\langle \bullet \rangle_0$ denotes the Gaussian expectation value in terms of the distribution $e^{-\mathcal{S_0}/\hbar}/\mathcal{Z}_0(\beta)$ and periodic boundary conditions. Here $\mathcal{Z}_0(\beta)$ is the partition function of the harmonic oscillator as 
\begin{equation}
\mathcal{Z}_0(\beta) = \frac{1}{2\sinh(\beta\omega/2)} = \frac{e^{-\beta\hbar\omega/2}}{1-e^{-\beta\hbar\omega}}.
\end{equation}
Then two-point correlations functions are given as
\begin{align}
\langle q(t_1)q(t_2) \rangle 
&= \mathcal{Z}_0^{-1}(\beta)\hbar^2 \left.\frac{\delta^2}{\delta b(t) \delta b(u)}\mathcal{Z}_G(\beta, b)\right|_{b=0} \nonumber \\
&= \frac{\hbar}{2\omega \tanh(\omega\tau/2)}.
\end{align}
It is no surprise that the temporal correlations are distinct from each other in this example. 
\begin{claim}
In general, temporal correlations in path integrals do not have the operational meaning as those in pseudo-density matrices since they use different measures, with exception of path-integral representation for spacetime states and decoherence functionals as Eqn.~\eqref{pich}.
\end{claim}
That indicates a fundamental difference of temporal correlations in path integrals and other spacetime approaches, and raises again the question whether probability or amplitude serves as the measure in quantum theory. 
It is natural that amplitudes interferes with each other in field theory and expectation values of operators are defined with amplitudes interference. Thus temporal correlations in path integrals cannot be operationally measured as pseudo-density matrices. 
However, spacetime states defined via position measurements and weak measurements in pseudo-density matrix formulation~\cite{zhang2020different} are motivated by the path integral formalism and have path-integral representations naturally. 
In addition, consistent histories also have a path-integral representation of decoherence functionals as we mentioned earlier in Eqn.~\eqref{pich}: 
\begin{align}
D([\alpha],& [\alpha']) = \int_{[\alpha]} \mathcal{D} q^i \int_{[\alpha']} \mathcal{D} q^{i'} \nonumber \\
&\exp(iS[q^i] - iS[q^{i'}]) \delta(q_f^i - q_f^{i'}) \rho(q_0^i, q_0^{i'}). \nonumber
\end{align}
(In the above we use Euclidean path integral for statistical mechanics, now we change to the usual convention.) 
Note that in consistent histories, the consistence conditions lead to the vanishing of the action part; that is, $D([\alpha], [\alpha'])= 0$ if $[\alpha] \neq [\alpha']$, and $D([\alpha], [\alpha]) = \int_{[\alpha]} \mathcal{D} q^i \rho(q_0^i, q_0^{i})$.
Thus, the path integral representation of consistent histories does not distinguish the difference between amplitudes and probabilities, and serve as a coordinator for two representations.

\section{Conclusion}
Via the pseudo-density matrix
formalism, we found several relations among the spacetime formulations of indefinite
causal structures, consistent histories, generalised nonlocal games, out-of-time-order correlation functions, and path integrals: 
(1) It is possible to map a process matrix to a
corresponding pseudo-density matrix under correlations in three different ways:
one-lab to one-event direct map, one-lab to one-event with double Hilbert spaces map, and one-lab to two-event map. \\
(2) The decoherence functional in consistent histories is the probabilities in temporal correlations of pseudo-density matrices. \\
(3) The probability in a quantum-classical signalling game with a trivial input at the later time corresponds to the probability in a pseudo-density matrix with quantum channels as measurements. \\
(4) OTOCs can be represented as temporal correlations in pseudo-density matrices with half the numbers of steps for calculation; for example, a four-point OTOC, usually calculated by evolving forwards and backwards twice, is represented by a tripartite pseudo-density matrix with only once evolving forwards and backwards. \\
(5) In general, temporal correlations in path integrals
do not have the operational meaning as those in pseudo-density matrices since
they use different measures, with the exception of the path-integral
representation for spacetime states and decoherence functionals. \\
We conclude that under this comparison of temporal correlations the different approaches, except path integrals, are closely related. The path integral approach of quantum mechanics defines temporal correlation differently, weighted by amplitudes rather than probabilities. We hope that these relations can further aid a flow of ideas between the different approaches. How to move on to relativistic quantum information, or further to quantum gravity, is still a big gap worth exploring.

\acknowledgments
T.Z. thanks Lucien Hardy, Giulio Chiribella, Kavan Modi, Fabio Costa, and David Felce for discussion on indefinite causal structures; thanks Seth Lloyd, Robert Spekkens, and Wojciech Zurek for introducing consistent histories and decoherence functional; thanks Francesco Buscemi and Denis Rosset for discussion on quantum-classical signalling games; thanks Beni Yoshida, Nick Hunter-Jones, and Zi-Wen Liu for discussion on OTOCs; thanks Rafael Sorkin for discussion on path integrals and quantum measure; and thanks Lee Smolin for general discussion on time and correlations. T.Z. thanks Mateus Araújo for helpful comments on the first draft. T.Z. thanks the visiting graduate fellow program at Perimeter Institute for Theoretical Physics. O.D. acknowledges National Natural Science Foundation of China (NSFC). V.V. acknowledges funding from the National Research Foundation (Singapore), the Ministry of Education (Singapore), the Engineering and Physical Sciences Research Council (UK), and Wolfson College, University of Oxford.

\bibliography{refjiang}

\end{document}